\documentclass[11pt]{article}
\usepackage{color,soul}
\usepackage{titlesec}
\usepackage{breakcites}
\usepackage[pagebackref]{hyperref}
\usepackage{dirtytalk}
\usepackage{lipsum}
\usepackage{ntheorem}
\usepackage{mathtools}
\usepackage{graphicx}
\usepackage{calc}
\usepackage{stmaryrd}
\usepackage{paralist}
\newlength{\depthofsumsign}
\usepackage{graphicx}
\usepackage{amsmath}
\usepackage{amssymb}
\usepackage{amsfonts}
\usepackage{array}
\usepackage{tikz}
\usepackage{hhline}
\usepackage{multirow}
\usepackage{subfig}
\usepackage{hyperref}
\usepackage{mathtools}
\usepackage{tikz}
\usepackage{makecell}
\usepackage{pgfplots}
\usepackage{afterpage}
\usepackage{fancyhdr}
\usepackage{setspace}
\usepackage{bm}
\usepackage[linesnumbered,ruled,vlined]{algorithm2e}
\SetKwInput{KwInput}{Input}
\SetKwInput{KwOutput}{Output}
\usepackage{authblk}

\DeclarePairedDelimiter\floor{\lfloor}{\rfloor}

\newtheorem{proposition} {Proposition}
\newtheorem*{proposition-non}{Proposition}

\newtheorem{theorem}{Theorem}

\newtheorem{remark}{Remark}

\allowdisplaybreaks

\newenvironment{proof}{\noindent{\bf Proof:}\indent}%
                      {\hfill $\Box$\par}

\newcommand{\sym}[1]{{\sf #1}}

\title{The Nonlinear Filter Model of Stream Cipher Redivivus}
\author[1,2]{Claude Carlet}
\author[3]{Palash Sarkar\thanks{Corresponding author.}}
\affil[1]{LAGA Laboratory, University of Paris 8, 93526 Saint-Denis, France}
\affil[2]{University of Bergen, Norway}
\affil[3]{Indian Statistical Institute, 203, B.T. Road, Kolkata, India 700108}
\affil[ ]{Emails: {\tt claude.carlet@gmail.com, palash@isical.ac.in}}

\date{\today}

\begin{document}

\maketitle



%

\begin{abstract}
	The nonlinear filter model is an old and well understood approach to the design of secure stream ciphers. Extensive research over several decades has shown how to attack 
	stream ciphers based on this model and has identified the required security properties of the Boolean function used as the filtering function to resist such attacks. This 
	led to the problem of constructing Boolean functions which provide adequate security \textit{and} at the same time are efficient to implement. Unfortunately, over the last 
	two decades no fully satisfactory solutions to this problem appeared in the literature. The lack of good solutions has effectively led to the nonlinear filter model becoming 
	more or less 
	obsolete. This is a big loss to the cryptographic design toolkit, since the great advantages of the nonlinear filter model are its simplicity, well understood security and 
	the potential to provide low cost solutions for hardware oriented stream ciphers. In this paper, we revive the nonlinear filter model by constructing appropriate Boolean 
	functions which provide required security and are also efficient to implement. We put forward concrete suggestions of stream ciphers which are $\kappa$-bit secure against 
	known types of attacks for $\kappa=80$, 128, 160, 192, 224 and 256. For the 80-bit and the 128-bit security levels, the gate count estimates of our proposals compare quite 
	well to the famous ciphers Trivium and Grain-128a respectively, while for the 256-bit security level, we do not know of any other stream cipher design which has such a 
	low gate count. 
\end{abstract}

\section{Introduction}
The nonlinear filter model of stream ciphers is several decades old; one may note that the model was extensively discussed in the book by Rueppel~\cite{Ru86} which was 
published in the mid-1980s.
The nonlinear filter model consists of two components, namely a linear feedback shift register (LFSR) and a Boolean function $f$ which is applied
to a subset of the bits of the LFSR at fixed positions (called tap positions). 
At each clock cycle, $f$ is applied to the present state of the LFSR to produce a single keystream bit and simultaneously the LFSR also moves 
to the next state. LFSRs are efficient to implement in hardware. So the implementation efficiency of the nonlinear filter model is essentially determined 
by the efficiency of implementing $f$. 

A basic requirement on the filtering function $f$ is that it is balanced, as otherwise the keystream sequence produced by the stream cipher is unbalanced, which leads to a 
distinguishing attack. 
Extensive research carried out over the last few decades has shown several approaches to cryptanalysing the nonlinear filter model of stream ciphers. The initial line of 
attack was based upon determining
the linear complexity of the produced keystream. For an LFSR of length $L$ and a Boolean function of algebraic degree $d$, under certain reasonable and easy-to-ensure
conditions, the linear complexity of the keystream is known to be at least ${L\choose d}$ (see~\cite{Ru86}). Using large enough values of $L$ and $d$, linear complexity
based attacks can be made infeasible.
The second phase of attacks consisted of various kinds of (fast) correlation attacks. Starting with the first such attack in 
1985~\cite{DBLP:journals/tc/Siegenthaler85} (which was efficient on another model, the nonlinear combiner model), a long line of 
papers~\cite{DBLP:conf/eurocrypt/Siegenthaler85,DBLP:journals/joc/MeierS89,DBLP:conf/eurocrypt/GolicP92,DBLP:conf/fse/Anderson94,DBLP:conf/fse/ClarkGD96,DBLP:conf/crypto/JohanssonJ99,DBLP:conf/eurocrypt/JohanssonJ99,DBLP:conf/fse/ChepyzhovJS00,DBLP:conf/eurocrypt/CanteautT00,DBLP:journals/ipl/JonssonJ02,CF02,DBLP:conf/sacrypt/LeveillerZGB02,DBLP:conf/eurocrypt/ChoseJM02,DBLP:conf/crypto/TodoIMAZ18,DBLP:journals/tosc/ZhouFZ22,DBLP:journals/tosc/ZhangLGJ23,DBLP:journals/tit/MaJGCS24,DBLP:journals/dcc/MartinezS24} explored various avenues 
for mounting correlation attacks. 
Surveys of some of the older attacks appear in~\cite{Can05,DBLP:reference/crypt/Canteaut11d,DBLP:conf/fse/Meier11,DBLP:journals/ccds/AgrenLHJ12}.
Fast correlation attacks are based on affine approximation and apply to the nonlinear filter (as well as the nonlinear combiner) model.
The resistance to fast correlation attacks is mainly determined by the linear bias of the Boolean function $f$. The linear bias is determined by the nonlinearity 
of the function $f$; the higher the nonlinearity, the lower the linear bias.
The third phase of attacks started in 2003 with the publication of the algebraic attack~\cite{DBLP:conf/eurocrypt/CourtoisM03} and was soon followed
by the publication of the fast algebraic attack~\cite{DBLP:conf/crypto/Courtois03}. Resistance to these attacks requires the function $f$ to possess high
(fast) algebraic immunity.

The various attacks mentioned above have posed the following design challenge for a Boolean function to be used in the nonlinear filter model of stream ciphers. 
Construct balanced Boolean functions which achieve a good combination of nonlinearity and algebraic resistance \textit{and} are also very efficient to implement.
Unfortunately, since the time algebraic attacks were proposed in the early-2000s, no good solutions to the design challenge for Boolean functions have appeared in the literature
(some Boolean functions satisfy all the necessary cryptographic criteria, but are too heavy and slow to compute~\cite{DBLP:conf/asiacrypt/CarletF08}, and 
some others are fast to compute but possess insufficient nonlinearity~\cite{WCST,DBLP:journals/amco/WangTS14,DBLP:journals/tit/Carlet22,MO24}).
A consequence of not being able to find good solutions to the design challenge is that the nonlinear filter model of stream ciphers became obsolete. This is somewhat 
unfortunate since the model being very old, it is well studied with well understood security, and has the potential to provide low gate count solutions in hardware.
We note that a recent independent work\footnote{We note that the preprint version of the present work appeared before~\cite{DBLP:journals/tosc/ChacalGMMP25}.} has put 
forward a concrete proposal of the nonlinear filter model of stream ciphers tailored for hybrid homomorphic encryption.~\cite{DBLP:journals/tosc/ChacalGMMP25}.

A class of guess-then-determine attacks is known against the nonlinear filter model. The inversion attack~\cite{DBLP:conf/fse/Golic96} is the first such attack.
Subsequently, a number of papers have developed the idea into the generalised inversion attack, the filter state guessing attack, and the generalised filter state guessing 
attack~\cite{DBLP:conf/fse/Golic96,DBLP:journals/tc/GolicCD00,DBLP:journals/iet-ifs/HodzicPW19,DBLP:conf/balkancryptsec/PasalicHBW14,DBLP:journals/tit/WeiPH12}.
Recommendations for resistance to the state guessing attack do not specify conditions on the filtering function. Rather, the recommendations specify
conditions on the tap positions, i.e. the LFSR positions where the filtering function inputs are drawn. 
Strictly following these recommendations with an LFSR of usual size (say about 256 flip-flops) requires taking the number of variables of the filtering function to be much 
lower than the desired security level. Later we elaborate on this point. 

We note that even though the nonlinear filter model became obsolete, the use of LFSRs in the design of stream ciphers has continued for both hardware and software oriented 
proposals (see for example~\cite{DBLP:conf/isit/Hell0MM06,sosemanuk,DBLP:journals/tosc/Ekdahl0MY19}).
Instead of using a Boolean function, such designs typically use a nonlinear finite state machine to filter the output of the LFSR. 
Some ciphers such as~\cite{DBLP:series/lncs/CanniereP08,DBLP:series/lncs/BabbageD08} have gone further and replaced the LFSR with one or more 
nonlinear feedback shift registers (NFSRs). 

In this paper, we revisit the above mentioned design problem for Boolean functions towards the goal of reviving the nonlinear filter model. 
Bent functions~\cite{DBLP:journals/jct/Rothaus76} are a very well studied class of Boolean functions. They exist for even number of variables and provide the
highest possible nonlinearity. 
However, their use in cryptography cannot be direct, since they are unbalanced, and even if we modify them into balanced functions (as Dobbertin~\cite{DBLP:conf/fse/Dobbertin94} 
proposed) that use is not clear, after the invention of fast algebraic attacks.
The introduction of Chapter~6 of~\cite{BF-book} summarises the state-of-the-art as follows: 
``we do not know an efficient construction using bent functions which would provide Boolean functions having all the necessary features for being used in stream ciphers.''
Note that a result from~\cite{DBLP:conf/cisc/WangJ10} seemed to imply that it was impossible to obtain a Boolean function having a good resistance to fast algebraic attacks 
by modifying a bent function into a balanced function. This result happened to be incorrect as shown in~\cite{BF-book} (see Theorem~22 which corrects the result, and 
the few lines before it). Even after this correction, the problem of building good cryptographic functions from bent functions seemed hard. In the present paper we provide a 
concrete solution to this problem. We even show that it is possible to start with a bent function obtained from the very basic Maiorana-McFarland construction.

The well known Maiorana-McFarland class of bent functions is defined as follows. For $m\geq 1$, let $\mathbf{X}$ and
$\mathbf{Y}$ be two vectors of $m$ variables. Then a $2m$-variable Maiorana-McFarland bent function is defined to map $(\mathbf{X},\mathbf{Y})$ to 
$\langle \pi(\mathbf{X}),\mathbf{Y}\rangle \oplus h(\mathbf{X})$, where $\pi$ is a bijection from $m$-bit strings to $m$-bit strings and 
$h$ is any $m$-variable Boolean function. It is well known that the nonlinearity of Maiorana-McFarland functions does not depend on the choice of $h$ nor on 
that of the permutation $\pi$. Bad choices of both $\pi$ and $h$ (for example, $\pi$ to be the identity permutation and $h$ to be a constant function) provide
functions whose algebraic immunity is low. On the bright side, we observe that it may be possible to improve the algebraic resistance by properly choosing 
$h$. Assume that $\pi$ is chosen to be an affine map, i.e. the coordinate functions of $\pi$ are affine functions of its input variables.
It is known~\cite{DBLP:journals/dcc/DalaiMS06} that the majority function possesses
the best possible algebraic immunity. Motivated by this fact, we choose $h$ to be the majority function on $m$ variables. We \textit{prove} that the resulting
bent function on $2m$ variables has algebraic immunity at least $\lceil m/2\rceil$, and hence fast algebraic immunity at least $1+\lceil m/2\rceil$. On the implementation 
aspect, it is known that the majority function can be implemented using $O(m)$ gates (see Theorem 4.1 of~\cite{We87}).
So the obtained bent function has maximum nonlinearity, sufficient (fast) algebraic immunity when the number of variables is large enough (to protect 
against fast algebraic attacks at a specified security level), and at the same time is quite efficient to implement.
Indeed, the (fast) algebraic immunity is not the maximum possible, but due to the implementation efficiency, it is possible to increase the number of variables
to achieve the desired level of algebraic resistance. 
A novelty of our work is \textit{the observation and the proof} that choosing $h$ to be the majority function improves
the algebraic immunity. While Maiorana-McFarland bent functions have been extensively studied in the literature, this simple observation has escaped the notice of 
earlier researchers. To the best of our knowledge, our result is the first construction of an infinite class of bent functions with a provable non-trivial lower bound on 
algebraic immunity.
We note that a specific example of an 8-variable bent function with algebraic immunity 4 (the maximum possible value) was given in~\cite{DBLP:journals/ipl/WangT12}.

One problem is that a bent function is not balanced. This problem is easily rectified by XORing a new variable to the bent function, obtaining a function on an odd number 
of variables. This modification requires only one extra XOR gate for implementation. In terms of security, the modification does not change the linear bias. We prove that 
the algebraic immunity of the modified function is at least the algebraic immunity of the bent function, and hence the fast algebraic immunity is at least one plus
the algebraic immunity of the bent function. So the algebraic resistance of the modified function is essentially the same as that of the bent function. 
The resistance to fast correlation attacks and fast algebraic attacks are ensured by the bent function
(on an appropriate number of variables). A positive aspect of XORing a new variable is that it prevents (see Theorem~2 of~\cite{DBLP:conf/fse/Golic96}) certain kinds of 
information leakage which was identified in~\cite{DBLP:conf/fse/Anderson94}. On the negative side, filtering functions of the form $W+g(\mathbf{Z})$ were shown to 
be susceptible to the inversion attack~\cite{DBLP:conf/fse/Golic96}. We prevent this attack by choosing the gap between the first and the last tap positions to be sufficiently large; 
this countermeasure was already proposed in~\cite{DBLP:conf/fse/Golic96}. More generally, we show that due to the fact that we use filtering functions on a large
number of variables, the various kinds of state guessing 
attacks~\cite{DBLP:journals/tc/GolicCD00,DBLP:journals/iet-ifs/HodzicPW19,DBLP:conf/balkancryptsec/PasalicHBW14,DBLP:journals/tit/WeiPH12} do not apply to our proposals.

The literature~\cite{DBLP:conf/fse/Golic96,DBLP:journals/tc/GolicCD00,DBLP:journals/iet-ifs/HodzicPW19,DBLP:conf/balkancryptsec/PasalicHBW14,DBLP:journals/tit/WeiPH12} 
provides guidance on the choice of tap positions, i.e. the positions of the LFSR which are to be tapped to provide input to the filtering function. One such
recommendation is to ensure that the tap positions form a full positive difference set. As we show later, this condition requires the length $L$ of the LFSR to be 
at least quadratic in the number $n$ of variables of the filtering function. In our proposals, the value of $n$ is close to the target security level $\kappa$. 
So if the tap positions are to be chosen to satisfy the full positive difference set condition, then $L$ will become too large, and the resulting stream cipher will be
of no practical interest. The recommendations in the literature on selection of tap positions are essentially sufficient (but not necessary) conditions to prevent state 
guessing attacks. For our proposals, we show that such attacks can be avoided even though we do not follow the recommendations on tap positions.

We propose to use filtering function on a large number of variables. On the other hand, due to efficiency considerations, it is not possible to make the length
of the LFSR too long. In particular, in all our concrete proposals, the ratio $n/L$ is close to half and is much higher than what was used in all previous proposals.
Using a high value of $n/L$ means that the tap positions are placed much closer together than in previous proposals. This creates possibility of an overlap
in the two sets of state bits determining two nearby keystream bits. Such overlap has the potential to cancel out terms and reduce the overall ``complexity''. 
Our choice of tap positions is motivated by the requirement of ensuring that such cancellations do not take place. The generation of a keystream bit requires 
a call to the majority function. We set one of the goals of choosing the tap positions to be to ensure that the inputs to the two calls to majority for generating two 
different keystream bits have only a small overlap. In particular, we \textit{prove} that for our choice of tap positions, there is no cancellation of terms of the majority
function corresponding to two keystream bits. We evolved this design criterion for the tap positions in response to an attack~\cite{cryptoeprint:2025/197} on an earlier 
version, and based on informal attack ideas (though not actual attacks) suggested by Subhadeep Banik, Willi Meier, and Bin Zhang also on a previous version. 
Further, we choose all the tap positions for the variables in $\mathbf{X}$ to the left of all the tap
positions for the variables in $\mathbf{Y}$. This choice combined with the choice of $\pi$ as the bit reversal permutation allows us to \textit{prove} that the
quadratic terms arising from the application of the filtering function to the state bits corresponding to two different states do not cancel out with each other. 
More generally we introduce a new idea, which we call the \textit{shift overlap minimisation} strategy, of selecting tap positions, the principle being to try and
minimise the maximum overlap that arises due to shifts.

We perform a detailed concrete security analysis of some of the well known attacks on the nonlinear filter model. As the outcome of this analysis, for various security
levels, we provide concrete proposals for stream ciphers based on the nonlinear filter model using the Boolean functions described above as the filtering
functions. A strong point in favour of these proposals is that at the appropriate security levels they provide \textit{provable} assurance against well known classes of attacks.
Further, we provide concrete gate count estimates for the entire circuit to implement the stream ciphers. 
For the 80-bit, 128-bit, 160-bit, 192-bit, 224-bit and the $256$-bit security levels, 
we propose using LFSRs of lengths 163, 257, 331, 389, 449, 521 and filtering functions on 75, 119, 143, 175, 203 and 231 variables respectively.
The gate count estimates for the 80-bit, 128-bit, 160-bit, 192-bit, 224-bit and the $256$-bit security levels are 
1797.5, 2840.5, 3596.0, 4265.5, 4942.5, and 5693.0  
NAND gates respectively. The gate count estimates for the 80-bit and the 128-bit security levels compare quite 
well\footnote{Following~\cite{DBLP:journals/ijwmc/AgrenHJM11} we estimated 8 NAND gates for a flip-flop, whereas Trivium estimated 12 NAND gates for
a flip-flop. Using 12 NAND gates for a flip-flop, our proposal at the 80-bit security level requires about 2464.5 NAND gates, whereas Trivium requires 
about 3488 NAND gates.} with famous ciphers such as Trivium~\cite{DBLP:series/lncs/CanniereP08} and Grain-128a~\cite{DBLP:journals/ijwmc/AgrenHJM11} which offer
80-bit and 128-bit security respectively. For the other security levels, we are not aware of other stream ciphers which have such low gate counts.
So our revival of the nonlinear filter model of stream ciphers leads to concrete proposals which offer a combination of both 
provable security against well known classes of attacks at a desired level of security and also low gate count.

We note that there are some old (prior to the advent of algebraic attacks) works~\cite{DBLP:conf/ches/SarkarM01,DBLP:journals/tc/SarkarM03,DBLP:conf/wisa/GuptaS04} on 
efficient implementation of Boolean functions on a large number of variables targeted towards the nonlinear combiner model of stream ciphers. 
There are also a few later works~\cite{DBLP:journals/jce/PasalicCZ17,DBLP:journals/iacr/ChattopadhyayMMRT23} on implementation of Maiorana-McFarland type functions.
These works, however, do not cover the implementation of the functions that we introduce, the reason being that these new functions themselves do not appear earlier
in the literature.

\paragraph{Comparison between the nonlinear filter model and some modern stream ciphers.}
Well known stream ciphers such as Trivium~\cite{DBLP:series/lncs/CanniereP08}, Mickey~\cite{DBLP:series/lncs/BabbageD08}, Grain-128a~\cite{DBLP:journals/ijwmc/AgrenHJM11}, 
SNOW~\cite{DBLP:journals/tosc/Ekdahl0MY19} and Sosemanuk~\cite{sosemanuk}, use novel and ingenious ideas. Nonetheless, these are standalone designs.
The nonlinear filter model, on the other hand, is a \textit{model} for stream cipher design. Due to the simplicity of the nonlinear filter model, provable properties of 
the filtering function
can be translated into provable resistance of the stream cipher against well known classes of attacks. In particular, for the proposals that we put forward,
the provable linear bias of the filtering 
function translates to provable protection against a large class of fast correlation attacks, and the provable lower bound on the (fast) algebraic immunity translates to provable 
resistance against (fast) algebraic attacks. 
Stream ciphers based on either nonlinear finite state machines, or nonlinear feedback shift registers do not enjoy this advantage, i.e. for such stream ciphers it is
very hard to obtain provable guarantees against various well known types of attacks. As an example, our proposal at the 128-bit security level generates keystream for
which the best linear approximation has \textit{provable} linear bias of $2^{-60}$, while at the same security level, for Sosemanuk~\cite{sosemanuk} the best 
known~\cite{DBLP:journals/tit/MaJGCS24} linear approximation has correlation 
$2^{-20.84}$, and it is not known whether there are approximations with higher correlations. 
Similarly, the time complexities of fast algebraic attacks against Trivium, Mickey, Grain-128a, SNOW and Sosemanuk are not known. It is believed that
these stream ciphers can withstand fast algebraic attacks. In contrast, we provide strong security guarantee that at the appropriate security levels, 
the fast algebraic attack is ineffective against the new nonlinear filter model based stream cipher proposals that we put forward.

One advantage of using LFSRs is that it is possible to provably ensure that the LFSR has a maximum period. For stream ciphers based on NFSRs, such 
as Trivium and Mickey, such provable assurance is not available. 
For stream ciphers which use a combination of LFSR and NFSR such as Grain-128a, it is possible to mount a divide-and-conquer attack. For example the attack 
in~\cite{DBLP:conf/crypto/TodoIMAZ18} on Grain-128a finds the state of the LFSR independently of the state of the NFSR. 
(In some ways this is reminiscent of the attack by Siegenthaler~\cite{DBLP:journals/tc/Siegenthaler85} on the nonlinear combiner model.) So even though Grain-128a
uses a 256-bit state, due to the divide-and-conquer strategy the full protection of the large state is not achieved. On the other hand, for the nonlinear filter model,
there is no known way to mount a divide-and-conquer attack. Our proposal at the 128-bit security level uses a 257-bit LFSR, and there is no known way to estimate half the 
state of the LFSR without involving the other half. 

Lastly, we note that the nonlinear filter model provides a scalable design, while it is not clear how to scale the ideas behind standalone designs such as those
in~\cite{DBLP:series/lncs/CanniereP08,DBLP:series/lncs/BabbageD08,DBLP:journals/ijwmc/AgrenHJM11,DBLP:journals/tosc/Ekdahl0MY19,sosemanuk}.
By properly choosing the LFSR and the filtering function, the nonlinear filter model can be instantiated to various security levels. This provides a 
\textit{family} of stream ciphers rather than a single stream cipher. The scalability of the design makes it easier to target different security levels and also 
to ramp up parameters in response to improvements of known attacks. For example, at the 128-bit security level, the complexity of the fast algebraic attack
on the proposal that we put forward is more than $2^{130.12}$, and the correlation of the best linear approximation of the keystream is $2^{-60}$. By increasing
the gate count by about 47 gates, it is possible to ensure that the complexity of the fast algebraic attack is more than $2^{135.83}$ and the best linear approximation
of the keystream is $2^{-64}$. 



\paragraph{Important note.} An earlier version of the proposal was attacked~\cite{cryptoeprint:2025/197} using a differential attack. In response, we have modified the 
proposal to resist the attack in~\cite{cryptoeprint:2025/197}. Though the modified proposal described in this version of the paper resists the attack 
in~\cite{cryptoeprint:2025/197}, we do not have any proof that the present proposal resists all kinds of differential attacks. 
We note that while provable properties of the filtering function can be translated to provable resistance against certain classes of attacks,
it is by no means true that these properties provide resistance against \textit{all} classes of attacks. 

\paragraph{Reference implementation.} We welcome further study of the concrete proposals that we have put forward. To help in such study, we have made a 
reference implementation of the proposals. The simple C code for the reference implementation can be obtained from the following link.
\begin{center}
	\url{https://github.com/palsarkar/nonlinear-filter-stream-cipher}
\end{center}
The DOI for the code is \url{https://doi.org/10.5281/zenodo.20442452}.

The paper is organised as follows. In Section~\ref{sec-prelim} we describe the preliminaries. 
The Boolean function construction is described in Section~\ref{sec-cons}. Section~\ref{sec-crypto} performs the concrete security analysis and
Section~\ref{sec-MM-eff} provides the gate count estimates. Finally, Section~\ref{sec-conclu} concludes the paper. 


\section{Preliminaries \label{sec-prelim}}
This section provides the notation and the basic definitions. For details on Boolean functions we refer to~\cite{BF-book}.

By $\#S$ we will denote the cardinality of a finite set $S$. The finite field of two elements will be denoted by $\mathbb{F}_2$, and for a positive integer $n$,
$\mathbb{F}_2^n$ will denote the vector space of dimension $n$ over $\mathbb{F}_2$. By $\oplus$, we will denote the addition operator over both $\mathbb{F}_2$ and 
$\mathbb{F}_2^n$. An element of $\mathbb{F}_2^n$ will be considered to be an $n$-bit binary string.

For $n\geq 0$, let $\mathbf{x}=(x_1,\ldots,x_n)$ be an $n$-bit binary string. The support of $\mathbf{x}$ is $\sym{supp}(\mathbf{x})=\{1\leq i\leq n: x_i=1\}$,
and the weight of $\mathbf{x}$ is $\sym{wt}(\mathbf{x})=\#\sym{supp}(\mathbf{x})$. 
By $\mathbf{0}_n$ and $\mathbf{1}_n$ we will denote the all-zero and all-one strings of length $n$ respectively. 
Let $\mathbf{x}=(x_1,\ldots,x_n)$ and $\mathbf{y}=(y_1,\ldots,y_n)$ be two $n$-bit strings. The distance between $\mathbf{x}$ and $\mathbf{y}$
is $d(\mathbf{x},\mathbf{y})=\#\{i: x_i\neq y_i\}$; the inner product of $\mathbf{x}$ and $\mathbf{y}$ is 
$\langle \mathbf{x},\mathbf{y}\rangle = x_1y_1 \oplus \cdots \oplus x_ny_n$; and we define $\mathbf{x}\leq \mathbf{y}$ if $x_i=1$ implies $y_i=1$ for 
$i=1,\ldots,n$. 

An $n$-variable Boolean function $f$ is a map $f:\mathbb{F}_2^n\rightarrow \mathbb{F}_2$. 
The weight of $f$ is $\sym{wt}(f)=\#\{\mathbf{x}\in\mathbb{F}_2^n:f(\mathbf{x})=1\}$; $f$ is said to be \textit{balanced} if $\sym{wt}(f)=2^{n-1}$. 

\paragraph{Algebraic normal form.}
Let $f$ be an $n$-variable function. The \textit{algebraic normal form (ANF) representation} of $f$ is the following:
$f(X_1,\ldots,X_n) = \bigoplus_{\bm{\alpha}\in\mathbb{F}_2^n} a_{\bm{\alpha}} \mathbf{X}^{\bm{\alpha}}$, where 
$\mathbf{X}=(X_1,\ldots,X_n)$; for $\bm{\alpha}=(\alpha_1,\ldots,\alpha_n)\in \mathbb{F}_2^n$, $\mathbf{X}^{\bm{\alpha}}$ denotes the monomial
$X_1^{\alpha_1}\cdots X_n^{\alpha_n}$; and $a_{\bm{\alpha}} \in \mathbb{F}_2$.
If $f$ is not the zero function, then the (algebraic) degree of $f$ is $\sym{deg}(f)=\max\{\sym{wt}(\bm{\alpha}):a_{\bm{\alpha}}=1\}$. By convention, 
we assume that the degree of the zero function is 0. Functions of degree at most 1 are said to be
affine functions. Affine functions with $a_{\mathbf{0}_n}=0$ are said to be linear functions. 

We have the following relations between the coefficients $a_{\bm{\alpha}}$ in the ANF of $f$ and the values of $f$ (see for example Section~2.2 of~\cite{BF-book}).
For $\mathbf{x},\bm{\alpha}\in\mathbb{F}_2^n$,
\begin{eqnarray}\label{eqn-ANF-TT}
	f(\mathbf{x}) = \bigoplus_{\bm{\beta}\leq \mathbf{x}} a_{\bm{\beta}} & \mbox{and} &  a_{\bm{\alpha}} = \bigoplus_{\mathbf{z}\leq \bm{\alpha}} f(\mathbf{z}).
\end{eqnarray}

\paragraph{Nonlinearity and Walsh transform.}
For two $n$-variable functions $f$ and $g$, the distance between them is $d(f,g)=\#\{\mathbf{x}\in\mathbb{F}_2^n:f(\mathbf{x})\neq g(\mathbf{x})\}$.
The \textit{nonlinearity} of an $n$-variable function $f$ is $\sym{nl}(f) = \min d(f,g)$, where the minimum is over all $n$-variable affine functions $g$. 
We define the linear bias of an $n$-variable Boolean function $f$ to be $\sym{LB}(f)=1/2-\sym{nl}(f)/2^n$. 

The Walsh transform of an $n$-variable function $f$ is the map $W_f:\mathbb{F}_2^n\rightarrow \mathbb{Z}$, where for $\bm{\alpha}\in\mathbb{F}_2^n$,
$W_f(\bm{\alpha}) = \sum_{\mathbf{x}\in\mathbb{F}_2^n} (-1)^{f(\mathbf{x}) \oplus \langle \bm{\alpha}, \mathbf{x} \rangle}$. 
The nonlinearity of a function $f$ is given by its Walsh transform as follows:
	$\sym{nl}(f) = 2^{n-1} - \frac{1}{2}\max_{\bm{\alpha} \in \mathbb{F}_2^n} |W_f(\bm{\alpha})|$.

A function $f$ such that $W_f(\bm{\alpha})=\pm 2^{n/2}$ for all $\bm{\alpha}\in\mathbb{F}_2^n$ is said to be a bent function~\cite{DBLP:journals/jct/Rothaus76}. 
Clearly such functions can exist only if 
$n$ is even. The nonlinearity of an $n$-variable bent function is $2^{n-1} - 2^{n/2-1}$ and this is the maximum nonlinearity that can be attained by $n$-variable functions.

A function is said to be plateaued if its Walsh transform takes only the values $0,\pm v$, for non-zero $v$.

\paragraph{Algebraic resistance.}
Let $f$ be an $n$-variable function. The \textit{algebraic immunity} of $f$ is defined~\cite{DBLP:conf/eurocrypt/CourtoisM03,DBLP:conf/eurocrypt/MeierPC04} as follows:
$\sym{AI}(f)=\min_{g\neq 0} \{\sym{deg}(g): \mbox{ either } gf=0, \mbox{ or }  g(f\oplus 1)=0\}$.
From the restatement of Theorem~6.0.1 of~\cite{DBLP:conf/eurocrypt/CourtoisM03} in the extended version of~\cite{DBLP:conf/eurocrypt/CourtoisM03} available
at \url{http://ntcourtois.free.fr/toyolili.pdf} (accessed on December 20, 2025), it follows that for any $n$-variable function
$f$, there is a function $g$ of degree at most $\lceil n/2\rceil$ such that $h=fg$ is of degree at most $\lfloor n/2\rfloor$. From this result it easily follows
that for any $n$-variable function $f$, $\sym{AI}(f)\leq \lceil n/2\rceil$. 
(One may observe that if $g=h$, then $h=fg$ implies $g(1\oplus f)=0$, and so $1\oplus f$ has an annihilator of degree at most $\lceil n/2\rceil$; and
if $g\neq h$, then from $h=fg$ we get $fh=f^2g=fg$ and so $f(g\oplus h)=0$ which implies that $f$ has an annihilator of degree at most $\lceil n/2\rceil$).

The fast algebraic attack (FAA) was introduced in~\cite{DBLP:conf/crypto/Courtois03}. The idea of the attack is based on the following observation.
Let $f$ be an $n$-variable function and suppose $g$ is another $n$-variable function of degree $e$ such that $gf$ has degree $d$. If both $e$ and $d$ are small,
then $f$ is susceptible to an FAA. Given $f$, and for $e$ and $d$ satisfying $e+d\geq n$, it is known~\cite{DBLP:conf/crypto/Courtois03} that there exist functions $g$ 
and $h$ with $\sym{deg}(g)=e$ and $\sym{deg}(h)\leq d$ such that $gf=h$. 
For any pair of functions $g$ and $h$ of degrees $e$ and $d$ respectively, satisfying $gf=h$, we have $h=gf=gf^2=(gf)f=hf$ and so if $h\neq 0$, then 
$h$ is an annihilator of $1\oplus f$ which implies $d\geq \sym{AI}(f)$. 
If $1\leq e<\sym{AI}(f)$, then $g$ is not an annihilator of $f$ and so $h=gf\neq 0$, and we obtain $e+d\geq 1+\sym{AI}(f)$.
The \textit{fast algebraic immunity (FAI)} of $f$ is a combined measure of resistance to both algebraic and fast algebraic attacks:
$\sym{FAI}(f)\allowbreak =\allowbreak \min \left( 2\sym{AI}(f), \allowbreak 
\min_{g\neq 0}\allowbreak \{\sym{deg}(g) \allowbreak + \allowbreak \sym{deg}(fg): 
\allowbreak 1 \allowbreak \leq \allowbreak \sym{deg}(g) \allowbreak < \allowbreak \sym{AI}(f)\}\right)$.
For any function $f$, $1+\sym{AI}(f)\leq \sym{FAI}(f)\leq 2\,\sym{AI}(f)$.

\paragraph{Majority function.}
For $n\geq 1$, let $\sym{Maj}_n:\{0,1\}^n\rightarrow \{0,1\}$ be the majority function defined in the following manner.
For $\mathbf{x}\in\{0,1\}^n$, $\sym{Maj}(\mathbf{x}) = 1$ if and only if $\sym{wt}(\mathbf{x}) > \floor{n/2}$.
Clearly $\sym{Maj}_n$ is a symmetric function. The following results were proved in~\cite{DBLP:journals/dcc/DalaiMS06}.
\begin{theorem}[Theorems~1 and~2 of~\cite{DBLP:journals/dcc/DalaiMS06}] \label{thm-maj-prop}
Let $n$ be a positive integer. 
	\begin{enumerate}
		\item $\sym{Maj}_n$ has the maximum possible AI of $\lceil n/2\rceil$.
		\item The degree of $\sym{Maj}_n$ is equal to $2^{\floor{\log_2n}}$.
		\item Any monomial occurring in the ANF of $\sym{Maj}_n$ has degree more than $\floor{n/2}$.
	\end{enumerate}
\end{theorem}


\section{Construction from Maiorana-McFarland Bent Functions \label{sec-cons}}
The Maiorana-McFarland class of bent functions is defined as follows. For $m\geq 1$, let $\pi:\{0,1\}^m\rightarrow\{0,1\}^m$ be a bijection and 
$h:\{0,1\}^m\rightarrow \{0,1\}$ be a Boolean function. Let $\pi_1,\ldots,\pi_m$ be the coordinate functions of $\pi$.
Let $\mathbf{X}=(X_1,\ldots,X_m)$ and $\mathbf{Y}=(Y_1,\ldots,Y_m)$. Given $h$ and $\pi$, for $m\geq 1$ the function $(h,\pi)\mbox{-}\sym{MM}_{2m}$ is defined to be 
the following.
\begin{eqnarray}
	(h,\pi)\mbox{-}\sym{MM}_{2m}(\mathbf{X},\mathbf{Y}) 
	& = & \langle \pi(\mathbf{X}),\mathbf{Y}\rangle \oplus h(\mathbf{X}) 
	= \pi_1(\mathbf{X})Y_1\oplus \cdots\oplus \pi_m(\mathbf{X})Y_m \oplus h(\mathbf{X}). \label{eqn-MM-even}
\end{eqnarray}
When the function $h$ and the permutation $\pi$ are clear from the context, we will simply write $\sym{MM}_{2m}$ instead of $(h,\pi)\mbox{-}\sym{MM}_{2m}$.

Since $\sym{MM}_{2m}$ is bent, $\sym{nl}(\sym{MM}_{2m}) = 2^{2m-1}-2^{m-1}$, and $\sym{LB}(\sym{MM}_{2m})=2^{-m-1}$.
Note that the nonlinearity of $\sym{MM}_n$ does not depend on the choices of the bijection $\pi$ and the function $h$. 
The degree of $\sym{MM}_{2m}$ is given by the following result.
\begin{proposition}\label{prop-MM-nl-deg} 
	For $m\geq 1$, $(h,\pi)\mbox{-}\sym{deg}(\sym{MM}_{2m})=\max(\sym{deg}(\pi_1)+1,\ldots,\sym{deg}(\pi_m)+1,\sym{deg}(h))$.
\end{proposition}

To the best of our knowledge the following result on the algebraic immunity of $\sym{MM}_{2m}$ is new.
\begin{theorem}\label{thm-MM-AI} 
	Suppose $m\geq 1$. There is an $\bm{\omega}^\star\in\mathbb{F}_2^m$ such that 
	\begin{eqnarray} \label{eqn-AI-MM-2m}
		\sym{AI}((h,\pi)\mbox{-}\sym{MM}_{2m}) & \geq & \sym{wt}(\bm{\omega}^\star) + \sym{AI}\big(\langle \bm{\omega}^\star, \pi(\mathbf{X})\rangle \oplus h(\mathbf{X})\big).
	\end{eqnarray}

	Suppose that $\pi$ is an affine map, i.e. for $1\leq i\leq m$, $\pi_i(X_1,\ldots,X_m)$ is an affine function. Then 
	\begin{eqnarray*}	
		\sym{AI}((h,\pi)\mbox{-}\sym{MM}_{2m}) & \geq  & \sym{AI}(h).
	\end{eqnarray*}	
\end{theorem}
\begin{proof}
	Suppose $g(\mathbf{X},\mathbf{Y})$ is an annihilator for $\sym{MM}_{2m}(\mathbf{X},\mathbf{Y})$. 
	Recall that for $\bm{\omega}=(\omega_1,\ldots,\omega_m)\in\mathbb{F}_2^m$, by $\mathbf{Y}^{\bm{\omega}}$ we denote the monomial 
	$Y_1^{\omega_1}\cdots Y_m^{\omega_m}$. Using this notation, we write 
	$g(\mathbf{X},\mathbf{Y})=\bigoplus_{\bm{\omega}\in\mathbb{F}_2^m} \mathbf{Y}^{\bm{\omega}}g_{\bm{\omega}}(\mathbf{X})$, for some functions 
	$g_{\bm{\omega}}(\mathbf{X})$'s. We have
	\begin{eqnarray}
		0 & = & g(\mathbf{X},\mathbf{Y}) \sym{MM}_{2m}(\mathbf{X},\mathbf{Y}) \nonumber \\
		& = & \left(\bigoplus_{\bm{\omega}\in\mathbb{F}_2^m} \mathbf{Y}^{\bm{\omega}}g_{\bm{\omega}}(\mathbf{X})\right)
		\big(\pi_1(\mathbf{X})Y_1\oplus \cdots\oplus \pi_m(\mathbf{X})Y_m \oplus h(\mathbf{X})\big). \label{eqn-tmp0}
	\end{eqnarray}
	Since the right hand side of~\eqref{eqn-tmp0} is equal to 0, for $\bm{\omega}\in\mathbb{F}_2^m$, the coefficient of $\mathbf{Y}^{\bm{\omega}}$
	in the expansion on the right hand side of~\eqref{eqn-tmp0} must be equal to 0.
	Since $g(\mathbf{X},\mathbf{Y})\neq 0$, let $w\geq 0$ be the minimum integer such that there is an $\bm{\omega}^\prime$ with $\sym{wt}(\bm{\omega}^\prime)=w$
	and $g_{\bm{\omega}^\prime}(\mathbf{X})\neq 0$. In~\eqref{eqn-tmp0}, equating the coefficient of $\mathbf{Y}^{\bm{\omega}^\prime}$ to 0, we have 
	\begin{eqnarray*}
		0 & = & g_{\bm{\omega}^\prime}(\mathbf{X})\left(h(\mathbf{X}) \oplus \left(\bigoplus_{i\in\sym{supp}(\bm{\omega}^\prime)}\pi_i(\mathbf{X})\right)\right) \\
		& = & g_{\bm{\omega}^\prime}(\mathbf{X}) \big(\langle \bm{\omega}^\prime, \pi(\mathbf{X})\rangle \oplus h(\mathbf{X})\big).
	\end{eqnarray*}
	So $g_{\bm{\omega}^\prime}(\mathbf{X})$ is an annihilator for $\langle \bm{\omega}^\prime, \pi(\mathbf{X})\rangle \oplus h(\mathbf{X})$. 
	Consequently, $\sym{deg}(g)\geq \sym{wt}(\bm{\omega}^\prime) + \sym{deg}(g_{\bm{\omega}^\prime}) 
	\geq \sym{wt}(\bm{\omega}^\prime) + \sym{AI}\big(\langle \bm{\omega}^\prime, \pi(\mathbf{X})\rangle \oplus h(\mathbf{X})\big)$.  

	Now suppose that $g(\mathbf{X},\mathbf{Y})$ is an annihilator for $1\oplus \sym{MM}_{2m}(\mathbf{X},\mathbf{Y})$. Then again writing 
	$g(\mathbf{X},\mathbf{Y})=\bigoplus_{\bm{\omega}\in\mathbb{F}_2^m} \mathbf{Y}^{\bm{\omega}}g_{\bm{\omega}}(\mathbf{X})$, an argument similar to the above
	shows that there is an $\omega^{\prime\prime}$ such that $g_{\bm{\omega}^{\prime\prime}}(\mathbf{X})$ is an annihilator for 
	$\langle \bm{\omega}^{\prime\prime}, \pi(\mathbf{X})\rangle \oplus 1\oplus h(\mathbf{X})$, and again we have
	$\sym{deg}(g) \geq \sym{wt}(\bm{\omega}^{\prime\prime}) + \sym{AI}\big(\langle \bm{\omega}^{\prime\prime}, \pi(\mathbf{X})\rangle \oplus h(\mathbf{X})\big)$.

	Combining the above two cases we get that there is a function $g$ which is either an annihilator of $\sym{MM}_{2m}(\mathbf{X},\mathbf{Y})$, or
	an annihilator of $1\oplus \sym{MM}_{2m}(\mathbf{X},\mathbf{Y})$, and there is some $\bm{\omega}^\star\in\mathbb{F}_2^n$ such that
	$\sym{deg}(g)\geq \sym{wt}(\bm{\omega}^\star) + \sym{AI}\big(\langle \bm{\omega}^\star, \pi(\mathbf{X})\oplus h(\mathbf{X})\rangle \big)$. 
	This shows~\eqref{eqn-AI-MM-2m}.

	Suppose now that $\pi$ is an affine map. Then $\ell(\mathbf{X})=\langle \bm{\omega}^\star, \pi(\mathbf{X})\rangle$ is an affine function. 
	If $w=0$, then $\langle \bm{\omega}^\star, \pi(\mathbf{X})\rangle \oplus h(\mathbf{X})=h(\mathbf{X})$, and so we have the result. So suppose $w>0$.
	From Lemma~1 of~\cite{DBLP:journals/tit/CarletDGM06}, we have $\sym{AI}(\ell(\mathbf{X})\oplus h(\mathbf{X}))\geq \sym{AI}(h(\mathbf{X}))-1$. So 
	\begin{eqnarray*}
		\sym{AI}(\sym{MM}_{2m}) 
		& \geq  & \sym{wt}(\bm{\omega}^\star) + \sym{AI}\big(\ell(\mathbf{X}) \oplus h(\mathbf{X})\big) \\
		& \geq & w + \sym{AI}(h(\mathbf{X}))-1 \geq \sym{AI}(h(\mathbf{X})).
	\end{eqnarray*} 
\end{proof}

\paragraph{\bf Extension to $(h,\pi)\mbox{-}\sym{MM}_n$, for odd $n$.}
We consider a folklore extension of $(h,\pi)\mbox{-}\sym{MM}_{2m}$ to odd number of variables. 
\begin{eqnarray}
	(h,\pi)\mbox{-}\sym{MM}_1(W) & = & W, \nonumber \\ 
	(h,\pi)\mbox{-}\sym{MM}_{2m+1}(W,\mathbf{X},\mathbf{Y}) & = & W\oplus (h,\pi)\mbox{-}\sym{MM}_{2m}(\mathbf{X},\mathbf{Y}), \quad \mbox{for } m\geq 1. \label{eqn-MM-odd}
\end{eqnarray}
Extending the case for even $n$, when the function $h$ and the permutation $\pi$ are clear from the context, for all $n\geq 1$, we will simply write $\sym{MM}_{n}$ instead of 
$(h,\pi)\mbox{-}\sym{MM}_{n}$.

The following result states the properties of $\sym{MM}_{2m+1}$.
\begin{proposition}\label{prop-MM-odd-nl-deg} For $m\geq 1$, $\sym{MM}_{2m+1}$ is balanced, and
	\begin{compactenum}
	\item $\sym{nl}(\sym{MM}_{2m+1})=2^{2m}-2^m$. In particular, $\sym{LB}(\sym{MM}_{2m+1})=\sym{LB}(\sym{MM}_{2m})=2^{-(m+1)}$.
		\item $\sym{deg}(\sym{MM}_{2m+1}) = \sym{deg}(\sym{MM}_{2m})$. 
		\item $\sym{AI}(\sym{MM}_{2m})\leq \sym{AI}(\sym{MM}_{2m+1})\leq 1 + \sym{AI}(\sym{MM}_{2m})$.
	\end{compactenum}
\end{proposition}
\begin{proof}
	The first point is well known. The second point is immediate from the definition of $\sym{MM}_{2m+1}$.

	For the third point, we prove a more general statement than what is required. We show that if $f(\mathbf{Z})$ is an $n$-variable function
	and $f_1(W,\mathbf{Z})=W\oplus f(\mathbf{Z})$, then $\sym{AI}(f)\leq \sym{AI}(h)\leq 1 + \sym{AI}(f)$.

	Clearly if $g(\mathbf{Z})$ is an annihilator for $f(\mathbf{Z})$ (resp. $1\oplus f(\mathbf{Z})$), then $(1\oplus W)g(\mathbf{Z})$ is an annihilator for
	$f_1(\mathbf{Z})$ (resp. $1\oplus f_1(\mathbf{Z})$). This shows the
	upper bound. Next we consider the lower bound. Suppose $g(W,\mathbf{Z})\neq 0$ is an annihilator for $f_1(W,\mathbf{Z})$.
	We write $g(W,\mathbf{Z})$ as $g(W,\mathbf{Z})=Wg_1(\mathbf{Z})+g_0(\mathbf{Z})$. Noting that $f_1(W,\mathbf{Z})=W\oplus f(\mathbf{Z})$, we obtain
	\begin{eqnarray*}
		0 & = & g(W,\mathbf{Z})f_1(W,\mathbf{Z}) \\
		& = & g_0(\mathbf{Z})f(\mathbf{Z}) \oplus W\big(g_0(\mathbf{Z}) \oplus g_1(\mathbf{Z})(1\oplus f(\mathbf{Z}))\big).
	\end{eqnarray*}
	So $g_0(\mathbf{Z})f(\mathbf{Z}) = 0$ and $g_0(\mathbf{Z}) \oplus g_1(\mathbf{Z})(1\oplus f(\mathbf{Z})) = 0.$
	If $g_0$ is non-zero, then $g_0$ is an annihilator for $f$ and so $\sym{deg}(g)\geq \sym{deg}(g_0)\geq \sym{AI}(f)$. If $g_0=0$, then
	since $g\neq 0$, it follows that $g_1\neq 0$. In this case, $g_1$ is an annihilator for $1\oplus f$, and so
	$\sym{deg}(g)=1+\sym{deg}(g_1)\geq 1+\sym{AI}(f)$. Consequently, in both cases $\sym{deg}(g)\geq \sym{AI}(f)$.

	On the other hand, if $g(W,\mathbf{Z})\neq 0$ is an annihilator for $1\oplus f_1(W,\mathbf{Z})$, then from $g(W,\mathbf{Z})(1\oplus f_1(W,\mathbf{Z}))=0$ and noting that
	$W(1\oplus W)=0$, we obtain $g_0(\mathbf{Z})(1\oplus f(\mathbf{Z}))=0$ and $g_0(\mathbf{Z}) \oplus g_1(\mathbf{Z})f(\mathbf{Z})=0$.
	If $g_0\neq 0$, then $g_0$ is an annihilator for $1\oplus f$, and if $g_0=0$, then $g_1$ is an annihilator for $f$.
	So again we have $\sym{deg}(g)\geq \sym{AI}(f)$. 

\end{proof}

From Theorem~\ref{thm-MM-AI}, choosing $\pi$ to be any affine permutation results in the AI of $(h,\pi)\mbox{-}\sym{MM}_{2m}$ to be lower bounded by the AI of $h$.
We make the following concrete choices.
\begin{quote}
	\textit{Concrete choice of $\pi$ in $(h,\pi)\mbox{-}\sym{MM}_{2m}$}: Choose the $m$-bit to $m$-bit permutation $\pi$ in the construction of $\sym{MM}_{2m}$ given by~\eqref{eqn-MM-even}
	to be the bit reversal permutation $\sym{rev}$, i.e. $\pi(X_1,\ldots,X_m)=\sym{rev}(X_1,\ldots,X_m)=(X_m,\ldots,X_1)$. 

	\textit{Concrete choice of $h$ in $(h,\pi)\mbox{-}\sym{MM}_{2m}$}: Choose the $m$-variable function $h$ in the construction of $\sym{MM}_{2m}$ given by~\eqref{eqn-MM-even} 
	to be $\sym{Maj}_m$, which is the $m$-variable majority function.
\end{quote}
Our choice of $\sym{Maj}_m$ for $h$ is motivated by the fact that the majority function has the maximum possible algebraic immunity (see Theorem~\ref{thm-maj-prop}).
There are other functions which achieve maximum algebraic immunity~\cite{DBLP:journals/tit/CarletDGM06} and these could also be chosen to instantiate $h$. Our
choice of $\sym{Maj}_m$ is the simplest choice of a function achieving maximum algebraic immunity.

The choice of $\pi$ as the bit reversal permutation instead of the identity permutation is to provide resistance to a type of differential attack which
was proposed on a previous version. The attack is discussed in details in Section~\ref{subsec-BV-attack}. See also Remark~\ref{rem-bit-perm}.

By $(\sym{Maj},\sym{rev})\mbox{-}\sym{MM}_{2m}$ we denote the function obtained by instantiating the definition of $\sym{MM}_{2m}$ given by~\eqref{eqn-MM-even}
with $h=\sym{Maj}_m$ and $\pi=\sym{rev}$. Similarly, by $(\sym{Maj},\sym{rev})\mbox{-}\sym{MM}_{2m+1}$, we denote the function obtained
from $(\sym{Maj},\sym{rev})\mbox{-}\sym{MM}_{2m}$ using~\eqref{eqn-MM-odd}.

\begin{proposition}\label{prop-deg-MM-sub-class}
	For $n\geq 4$, the degree of $(\sym{Maj},\sym{rev})\mbox{-}\sym{MM}_{n}$ is $2^{\floor{\log_2\floor{n/2}}}$. 
\end{proposition}
\begin{proof}
	The degree of $\sym{Maj}_m$ is $2^{\floor{\log_2m}}$ (see Theorem~\ref{thm-maj-prop}). Applying Proposition~\ref{prop-MM-nl-deg}, 
	we obtain the required result for even $n$. For odd $n$, the result follows from the second point of Proposition~\ref{prop-MM-odd-nl-deg}. 
\end{proof}

\begin{proposition}\label{prop-AI-n}
	For $n\geq 4$, if $n$ is even, then $\sym{AI}((\sym{Maj},\sym{rev})\mbox{-}\sym{MM}_{n})\geq \lceil n/4\rceil$, and if $n$ is odd
	then $\sym{AI}((\sym{Maj},\sym{rev})\mbox{-}\sym{MM}_{n})\geq \lceil (n-1)/4\rceil$.
\end{proposition}
\begin{proof}
	Suppose $n=2m$ is even. From Theorem~\ref{thm-MM-AI} we have $\sym{AI}((\sym{Maj},\sym{rev})\mbox{-}\sym{MM}_{2m})\geq \sym{AI}(\sym{Maj}_m)$. Since
	$\sym{AI}(\sym{Maj}_m)=\lceil m/2\rceil$, we have the result. For odd $n=2m+1$, from Proposition~\ref{prop-MM-odd-nl-deg}, we have 
	$\sym{AI}((\sym{Maj},\sym{rev})\mbox{-}\sym{MM}_{2m+1} \geq \sym{AI}((\sym{Maj},\sym{rev})\mbox{-}\sym{MM}_{2m}$, which shows the result. 
\end{proof} \ \\

From the results proved in~\cite{DBLP:journals/ipl/GuptaNG11,DBLP:journals/ipl/WangT12} it follows that 
$\sym{AI}((\sym{Maj},\sym{rev})\mbox{-}\sym{MM}_{n})\leq 2+ \lceil n/4\rceil$. So the lower bound given by Proposition~\ref{prop-AI-n} is close to the upper bound.
We computed the algebraic immunity of $(\sym{Maj},\sym{rev})\mbox{-}\sym{MM}_{n}$ for values of $n\leq 16$ and observed that for $n\geq 4$,
$\sym{AI}((\sym{Maj},\sym{rev})\mbox{-}\sym{MM}_{n})=1+\lfloor n/4 \rfloor$. This suggests that the lower bound in Proposition~\ref{prop-AI-n} is exact if $n\not\equiv 0\bmod 4$
and if $n\equiv 0\bmod 4$, the actual algebraic immunity is one more than the lower bound. 
In a work~\cite{DBLP:conf/latincrypt/Meaux25} subsequent to the present paper, it 
was shown that for even $n$, $\sym{AI}((\sym{Maj},\sym{rev})\mbox{-}\sym{MM}_{n})\leq 1+2^{\lceil\log(n/4)\rceil}$.
As mentioned in Section~\ref{sec-prelim}, for
any Boolean function $f$, $\sym{FAI}(f)\geq 1+\sym{AI}(f)$. Our experiments suggest that for $(\sym{Maj},\sym{rev})\mbox{-}\sym{MM}_{n}$,
the fast algebraic immunity is actually one more than the algebraic immunity. 

From the above discussion, it follows that the algebraic immunity of $(\sym{Maj},\sym{rev})\mbox{-}\sym{MM}_{n}$ is less than the maximum possible value $\lceil n/2\rceil$.
On the other hand, it can be implemented using $O(m)$ gates
(as we show in Section~\ref{sec-MM-eff}). So it is possible to increase the number of variables to achieve the desired level of algebraic resistance without increasing the
circuit size too much. 

The function $\sym{MM}_{2m+1}$ is balanced (see Proposition~\ref{prop-MM-odd-nl-deg}), and the cryptographic 
resistance provided by $\sym{MM}_{2m+1}$ is very similar to that provided by $\sym{MM}_{2m}$. In particular, the linear bias of
$\sym{MM}_{2m+1}$ is the same as that of $\sym{MM}_{2m}$, and the algebraic resistance of $\sym{MM}_{2m+1}$ is at least that of $\sym{MM}_{2m}$.
In terms of implementation efficiency, $\sym{MM}_{2m+1}$ requires only one extra XOR gate in addition to the circuit to implement $\sym{MM}_{2m}$.
The function $\sym{MM}_{2m+1}$ maps the all-zero string to 0. This can be a problem for use in a linear system, since such a system will map 
the all-zero string to the all-zero string.
We modify $\sym{MM}_{2m+1}$ so that the all-zero string is mapped to 1. We define $f_{2m+1}:\{0,1\}^{2m+1}\rightarrow \{0,1\}$ to be
\begin{eqnarray}\label{eqn-f}
	f_{2m+1}(W,\mathbf{X},\mathbf{Y}) & = & 1\oplus \sym{MM}_{2m+1}(W,\mathbf{X},\mathbf{Y}) 
	= 1\oplus W \oplus \langle \sym{rev}(\mathbf{X}),\mathbf{Y}\rangle \oplus \sym{Maj}_m(\mathbf{X}).
\end{eqnarray}
The cryptographic properties of $f_{2m+1}$ are exactly the same as those of $\sym{MM}_{2m+1}$. In the next section, we propose the use of $f_{2m+1}$ in the construction of the
filter generator model of stream ciphers.

\section{Concrete Stream Cipher Proposals \label{sec-crypto}}
We revisit the filter generator model where the state of an LFSR of length $L$ is filtered using a Boolean function. 
We refer to~\cite{LN96} for the general theory of LFSRs. 

In Figure~\ref{fig-schematic} we provide a schematic diagram of the nonlinear filter model of stream ciphers.
\textit{By $\mathcal{S}(L,m)$, we denote the class of stream ciphers obtained from an LFSR of length $L$ with a primitive connection polynomial, and with
$f_{2m+1}$ as defined in~\eqref{eqn-f} as the filtering function.}
For the $\kappa$-bit security level, we assume that the stream cipher supports a $\kappa$-bit secret key and a $\kappa$-bit initialisation vector (IV), and so $L\geq 2\kappa$. 
Below we describe the construction of $\mathcal{S}(L,m)$ which requires specifying a few more parameters in addition to $L$ and $m$.

\begin{figure}[htb]
\setlength{\unitlength}{4mm}
\begin{picture}(30,10)


	\put(8,4){\framebox(16,1){LFSR}}

	\put(16,4){\vector(0,-1){2}} \put(16.05,3){\makebox(0.25,0.25){{\tiny $\backslash r$}}}
	\put(16,2){\vector(-1,0){4}}
	\put(11.5,1.75){\makebox(0.5,0.5){$\oplus$}}
	\put(11.5,2){\vector(-1,0){6}}
	\put(5.5,2){\vector(0,1){2.5}}
	\put(5.5,4.5){\vector(1,0){2.5}}

	\put(23,7){\framebox(3,3){$f_{2m+1}$}} 
	\put(26,8.5){\vector(1,0){3}} \put(29,8.5){\makebox(1,1){\tiny o/p bit}}

	\put(10,5){\vector(0,1){4.5}} \put(10.15,6){\makebox(0.25,0.25){{\tiny $\backslash m$}}}
	\put(10,9.5){\vector(1,0){13}} \put(20,9.75){\makebox(0.25,0.25){{\tiny $\mathbf{X}$}}}

	\put(14,5){\vector(0,1){3.5}} \put(14.15,6){\makebox(0.25,0.25){{\tiny $\backslash m$}}}
	\put(14,8.5){\vector(1,0){9}} \put(20,8.75){\makebox(0.25,0.25){{\tiny $\mathbf{Y}$}}}

	\put(19,5){\vector(0,1){2.5}} \put(19.05,6){\makebox(0.25,0.25){{\tiny $\backslash 1$}}}
	\put(19,7.5){\vector(1,0){4}} \put(20,7.75){\makebox(0.25,0.25){{\tiny $W$}}}
\end{picture}
	\caption{Schematic diagram of keystream generation by the nonlinear filter model of stream ciphers. Here $r$ is the number of tap positions
	required to obtain the next bit of the LFSR. \label{fig-schematic} }
\end{figure}

\paragraph{LFSR maps.} Let $\tau(x)=x^L\oplus c_{L-1}x^{L-1}\oplus\cdots\oplus c_1x\oplus c_0$ be a primitive polynomial of degree $L$ over $\mathbb{F}_2$. 
Using $\tau$, we define the linear next bit map $\sym{nb}:\{0,1\}^L\rightarrow \{0,1\}$ and the linear next state map $\sym{NS}:\{0,1\}^L\rightarrow \{0,1\}^L$
as follows.
\begin{eqnarray}
	\sym{nb}(u_{L-1},\ldots,u_1,u_0) & = & c_{L-1}u_{L-1}\oplus \cdots \oplus c_1u_1\oplus c_0u_0, \label{eqn-next-bit} \\
	\sym{NS}(u_{L-1},\ldots,u_1,u_0) & = & (\sym{nb}(u_{L-1},\ldots,u_1,u_0),u_{L-1},\ldots,u_1). \label{eqn-next-st}
\end{eqnarray}
It is easy to note that the map $\sym{NS}$ is efficiently invertible. We define the vector $\mathbf{c}$ to be the following.
\begin{eqnarray}\label{eqn-LFSR-tap-pos}
	\mathbf{c} & = & (c_{L-1},\ldots,c_1,c_0).
\end{eqnarray}
The positions of $\mathbf{c}$ where the value is 1 are the positions of the LFSR that are tapped to obtain the next bit of the LFSR.

\paragraph{Filter function and tap positions.} The function $f_{2m+1}(W,\mathbf{X},\mathbf{Y})$ (given by~\eqref{eqn-f}), where $\mathbf{X}=(X_1,\ldots,X_m)$ and 
$\mathbf{Y}=(Y_1,\ldots,Y_m)$, is a 
function on $2m+1$ variables. This function will be used as the filtering function and applied to a subset of the bits of the $L$-bit state. We next describe the
tap positions of the $L$-bit state which provide the input bits to $f_{2m+1}$.

	In the concrete stream cipher proposals we put forward, $2m+1$ is less than the security parameter $\kappa$. Hence, 
\textit{in what follows we assume $2m+1<\kappa$.} Let $i_1,\ldots,i_m,j_1,\ldots,j_m$ be integers satisfying the following condition.
\begin{eqnarray}\label{eqn-tap-pos}
	0=i_1 < i_2 < \cdots < i_{m-1} < i_m <\kappa \leq j_1 < j_2 < \cdots < j_{m-1} < j_m=2\kappa-2.
\end{eqnarray}
Given the integers $i_1,\ldots,i_m$ and $j_1,\ldots,j_m$ we define a function $\sym{proj}:\{0,1\}^L\rightarrow \{0,1\}^{2m+1}$ in the following manner. 
\begin{eqnarray}\label{eqn-proj}
	\sym{proj}(u_{L-1},\ldots,u_0) & = & (u_{L-2\kappa},u_{L-1-i_1},\ldots,u_{L-1-i_m},u_{L-1-j_1},\ldots,u_{L-1-j_m}).
\end{eqnarray}
The function $\sym{proj}$ extracts the subset of $2m+1$ bits from the $L$-bit state to which the function $f_{2m+1}$ is to be applied.
We define the composition $f_{2m+1}\circ\sym{proj}$ to be the following $L$-variable Boolean function.
\begin{eqnarray}\label{eqn-filter}
	(u_{L-1},\ldots,u_0) & \mapsto & f_{2m+1}(\sym{proj}(u_{L-1},\ldots,u_0)).
\end{eqnarray}
Note that the tap position for $W$ is $L-2\kappa$, the tap position for $X_p$ is $L-1-i_p$, and the tap position for $Y_{p^\prime}$ is $L-1-j_{p^\prime}$,
where $1\leq p,p^\prime\leq m$. There is no tap position in the rightmost $L-2\kappa$ positions of the state; the leftmost position (i.e. position $L-1$) of the state 
is the tap position for $X_1$, all the tap positions for the variables in $\mathbf{X}$ are in the leftmost $\kappa$ positions of the state; all the tap positions for the
variables in $\mathbf{Y}$ are in the next $\kappa$ positions of the state, and the position $L-2\kappa+1$ of the state is the tap position for $Y_m$.
We have
\begin{eqnarray}
	\lefteqn{f_{2m+1}(\sym{proj}(u_{L-1},\ldots,u_0))} \nonumber \\
	& = & 1\oplus u_{L-2\kappa} \oplus \left( \bigoplus_{p=1}^m u_{L-1-i_p}u_{L-1-j_{m+1-p}} \right) \oplus \sym{Maj}_m(u_{L-1-i_1},\ldots,u_{L-1-i_m}) \nonumber \\
	& = & 1\oplus u_{L-2\kappa} 
	\oplus u_{L-1-i_m}u_{L-1-j_1}\oplus \cdots \oplus u_{L-1-i_1}u_{L-1-j_m} \oplus \sym{Maj}_m(u_{L-1-i_1},\ldots,u_{L-1-i_m}).
\end{eqnarray}
Let $\sym{pos}[L-1,\ldots,0]$ be an $L$-bit string such that 
\begin{tabbing}
	\ \ \ \ \= \kill
	\> $\sym{pos}[L-1-i_p]=1$, for $1\leq p\leq m$; \\
	\> $\sym{pos}[L-1-j_{p^\prime}]=1$, for $1\leq p^\prime\leq m$; \\
	\> $\sym{pos}[L-2\kappa]=1$; \\
	\> and $\sym{pos}$ is 0 at all the other $L-2m-1$ positions.
\end{tabbing}
Note that the indexing of $\sym{pos}$ has $L-1$ as the leftmost position and $0$ as the rightmost position.
The string $\sym{pos}$ encodes the $2m+1$ tap positions of the state. Let $\sym{posX}$ be the leftmost $\kappa$-bit string of $\sym{pos}$, i.e. the segment of $\sym{pos}$
given by $\sym{pos}[L-1,\ldots,L-\kappa]$. The string $\sym{posX}$ encodes the tap positions of the variables in $\mathbf{X}$.

\begin{remark}\label{rem-bit-perm}
	In the filtering function $f_{2m+1}$, the particular choice of the bit permutation to instantiate $\pi$ is to be seen in conjunction with the choice of the 
	tap positions for the variables $X_1,\ldots,X_m$. We have chosen $\pi$ to be the bit reversal permutation $\sym{rev}$. The tap positions for $X_1,\ldots,X_m$
	are chosen keeping $\sym{rev}$ in the mind. If instead of $\sym{rev}$, we choose some other bit permutation for $\pi$, then a corresponding 
	permutation has to be applied to the tap positions for $X_1,\ldots,X_m$.
\end{remark}

\paragraph{Initialisation phase.} The stream cipher uses a $\kappa$-bit key $(k_{\kappa-1},\ldots,k_0)$ and a $\kappa$-bit initialisation vector $(v_{\kappa-1},\ldots,v_0)$. 
The state of the stream cipher is the state of the LFSR which is an $L$-bit string. In the initialisation phase, the concatenation of the key and the IV is padded to obtain
an $L$-bit string which is loaded into the LFSR. Then the stream cipher is clocked for $2\kappa$ rounds. In these rounds the keystream bit that is generated is not
produced as output. Instead it is fed back into the state of the LFSR. The goal of the initialisation phase is to obtain a good mix of the key and the IV bits, i.e.
at the end of the initialisation phase each bit of the state is a ``complex'' nonlinear function of the key and the IV bits. We design the round function of the
initialisaton phase to be an invertible map so that the entire initialisation phase is invertible. This ensures that the entropy of the key remains unchanged during
the initialisation phase. Below we provide the details of the initialisation phase. 

First we define the initialisation round function. To do this we need an $L$-bit string $\mathbf{d} = (d_{L-1},\ldots,d_0)$. We put the following
conditions on $\mathbf{d}$. 
\begin{eqnarray}
\left.
	\begin{array}{rcll}
		d_{L-1} & = & 1, & \\
		    d_i & = & 0 & \mbox{for } i=L-2\kappa-1,\ldots,0, \\
		    d_i & = & 0 & \mbox{if either } c_{i+1}=1 \mbox{ or } \sym{pos}[i+1]=1,\mbox{ for } i=L-2,\ldots,L-2\kappa.
	\end{array}
	\right\} \label{eqn-feed-back-cond}
\end{eqnarray}
Given the string $\mathbf{d}$ we define the initialisation round function $\sym{IR}:\{0,1\}^L\rightarrow \{0,1\}^L$ as follows. For $\mathbf{u}=(u_{L-1},\ldots,u_0)$,
we define $\sym{IR}(\mathbf{u})$ to be the $L$-bit string $\mathbf{w}=(w_{L-1},\ldots,w_0)$, where
\begin{eqnarray}\label{eqn-IR}
	(w_{L-1},w_{L-2},\ldots,w_0) & = & (\sym{nb}(\mathbf{u}),u_{L-1},\ldots,u_1) \oplus (d_{L-1}b,d_{L-2}b,\ldots,d_0b),
\end{eqnarray}
with $b=f_{2m+1}(\sym{proj}(u_{L-1},\ldots,u_0))$. In other words, on input $\mathbf{u}$ the function $\sym{IR}$ clocks the LFSR once and feeds back the keystream bit
$b$ obtained from $\mathbf{u}$ to the positions of the state corresponding to the positions where the string $\mathbf{d}$ has the value 1. 
In~\eqref{eqn-feed-back-cond}, the first condition says that feedback is always provided to the leftmost bit of the state, the second condition says that no feedback is 
provided to the rightmost $L-2\kappa$ bits of the state, and the third condition says that if the $i$-th position is tapped either for the next state bit of the LFSR, or for 
the input to the function $f_{2m+1}$, then no feedback is provided to the $(i-1)$-st position of the state. This last condition on $\mathbf{d}$ is required in the next result 
which shows that the function $\sym{IR}$ is invertible.

\begin{proposition}\label{prop-IR-inv}
	If $\sym{pos}[0]=0$, then the function $\sym{IR}$ defined in~\eqref{eqn-IR} is invertible.
\end{proposition}
\begin{proof}
	Let $\mathbf{w}=\sym{IR}(\mathbf{u})$, with $\mathbf{u}=(u_{L-1},\ldots,u_0)$ and $\mathbf{w}=(w_{L-1},\ldots,w_0)$. Also, let 
	$b\allowbreak =\allowbreak f_{2m+1}(\sym{proj}(\allowbreak u_{L-1},\allowbreak \ldots,\allowbreak u_0))$. We argue that the string $\mathbf{w}$ uniquely determines 
	the string $\mathbf{u}$. The bits of $\mathbf{u}$ which are provided as input to $f_{2m+1}$ are determined by the positions where the string $\sym{pos}$ is equal to 1. 
	Since by hypothesis $\sym{pos}[0]=0$, 
	the bit $b$ does not depend on the bit $u_0$. We argue that the bits of $\mathbf{u}$ which determine $b$ can be determined from $\mathbf{w}$.

	From the conditions on $\mathbf{d}$ given in~\eqref{eqn-feed-back-cond}, we have $d_{L-1}=1$ and so
	from~\eqref{eqn-IR} we have $w_{L-1}=\sym{nb}(\mathbf{u})\oplus b=c_{L-1}u_{L-1}\oplus\cdots\oplus c_1u_1\oplus c_0u_0\oplus b
	=c_{L-1}u_{L-1}\oplus\cdots\oplus c_1u_1\oplus u_0\oplus b$, since $c_0=1$ due to the connection polynomial of the LFSR being primitive.
	So $u_0=w_{L-1}\oplus c_{L-1}u_{L-1}\oplus\cdots\oplus c_1u_1\oplus b=w_{L-1}\oplus c\oplus b$, where 
	$c=c_{L-1}u_{L-1}\oplus\cdots\oplus c_1u_1$. Further, for $i=L-1,\ldots,1$, $u_i=w_{i-1}\oplus d_{i-1}b$ and
	consequently, $u_i=w_{i-1}$ whenever $d_{i-1}=0$. From the conditions on $\mathbf{d}$ given in~\eqref{eqn-feed-back-cond}, $d_{i-1}=0$ if the
	$i$-th position is tapped either for computing $\sym{nb}(\mathbf{u})$, or for the input to the function $f_{2m+1}$. In other words, for any $i$ in $\{1,\ldots,L-1\}$,
	if $u_i$ is involved in either the computation of $\sym{nb}(\mathbf{u})$, or the computation of $b$, then $u_i$ is simply equal to $w_{i-1}$. 
	So given $\mathbf{w}$, both the bits $b$ and $c$ are uniquely defined. As a result, $u_0=w_{L-1}\oplus c\oplus b$ is also 
	uniquely defined, and so is $u_i=w_{i-1}\oplus d_{i-1}b$, for $i=L-1,\ldots,1$ and $d_{i-1}=1$.
\end{proof}
While Proposition~\ref{prop-IR-inv} states that $\sym{IR}$ is invertible, the proof actually shows that given $\mathbf{w}=\sym{IR}(\mathbf{u})$ it is easy to obtain $\mathbf{u}$, 
i.e. $\sym{IR}$ is efficiently invertible.

Intuitively, feeding back the nonlinear bit $b$ to a large number of positions (subject to the constraints which ensure invertibility of the round function) 
potentially leads to a more complex and nonlinear state. 
On the other hand, the implementation of~\eqref{eqn-IR} requires a XOR gate for each $i$ such that $d_i=1$, for a total of $\sym{wt}(\mathbf{d})$ XOR gates. 
So too many feedback positions increases the gate count. As a compromise between these two considerations we propose that the weight of $\mathbf{d}$ be taken to be
$\mu=\lfloor \sqrt{2\kappa}\rfloor$ and the positions of the 1's in $\mathbf{d}$ are more or less equi-spaced so that the difference between two 1-positions
is also about $\mu$. Later for the concrete stream cipher proposals, we provide the corresponding $\mathbf{d}$'s.

\begin{remark}\label{rem-IR-inv-cond}
	In all our concrete proposals provided later, we have $L>2\kappa$. So the set of indices $\{0,\ldots,L-2\kappa-1\}$ is non-empty and from the definition of the 
	tap positions for $f_{2m+1}$, we have
	$\sym{pos}[L-2\kappa-1]=\cdots=\sym{pos}[0]=0$. In particular, $\sym{pos}[0]=0$ and so the condition given in Proposition~\ref{prop-IR-inv} holds. As a result,
	all the instantiations of the function $\sym{IR}$ in our concrete proposals are invertible.
\end{remark}

Next we show how $\sym{IR}$ is used to perform the initialisation phase.
Let $(b_{L-2\kappa-1},\ldots,b_0)$ be a bit string of length $L-2\kappa$ defined as follows: $(b_{L-2\kappa-1},\ldots,b_0)$ is equal to $(01)^{(L-2\kappa)/2}$ if $L-2\kappa$ is 
even and $(b_{L-2\kappa-1},\ldots,b_0)$ is equal to $1(01)^{(L-2\kappa-1)/2}$ if $L-2\kappa$ is odd. 
The string $(b_{L-2\kappa-1},\ldots,b_0)$ is used to pad the concatenation of the key and IV to obtain a string of length $L$. 
The initial $L$-bit state $\mathbf{s}$ of the LFSR is
\begin{eqnarray}\label{eqn-init-st}
	\mathbf{s} & = & (k_{\kappa-1},\ldots,k_0,v_{\kappa-1},\ldots,v_0,b_{L-2\kappa-1},\ldots,b_0).
\end{eqnarray}
Our choice of the padding string is somewhat 
arbitrary (other than being an almost balanced string). In particular, we do not know whether there is any security implication of using some other string for padding.

The initialisation round function is applied to update the initial state $\mathbf{s}$ given in~\eqref{eqn-init-st} a total of $2\kappa$ times as follows.
\begin{tabbing}
	\ \ \ \ \=\ \ \ \ \= \kill
	\> set $\mathbf{s}$ as in~\eqref{eqn-init-st} \\
	\> for $i\leftarrow 1\mbox{ to } 2\kappa$ do \\
	\>\> $\mathbf{s}\leftarrow \sym{IR}(\mathbf{s})$ \\
	\> end for \\
	\> output $\mathbf{s}$
\end{tabbing}

\paragraph{Keystream generation phase.} 
The generation of the keystream bits $z_0,z_1,z_2,\ldots$ is done as follows. 
\begin{tabbing}
        \ \ \ \ \=\ \ \ \ \= \kill
	\> let $\mathbf{s}$ be the output of the initialisation phase \\
        \> for $t\geq 0$ do \\
	\>\> $z_t\leftarrow f_{2m+1}(\sym{proj}(\mathbf{s}))$ \\
        \>\> $\mathbf{s}\leftarrow \sym{NS}(\mathbf{s})$ \\
        \> end for.
\end{tabbing}
Suppose that the state of the stream cipher at the end of the initialisation phase is $\mathbf{s}^{(0)}=(s_{L-1},\ldots,s_0)$. For $t\geq 1$, define
$s_{L+t-1}=\sym{nb}(s_{L+t-2},\ldots,s_{t-1})$, and 
\begin{eqnarray}\label{eqn-state-t}
	\mathbf{s}^{(t)} & = & (s_{L+t-1},s_{L+t-2},\ldots,s_t). 
\end{eqnarray}
Then $z_t=f_{2m+1}(\sym{proj}(\mathbf{s}^{(t)}))$.

\paragraph{Length of keystream.} 
We stipulate that from a single key and IV pair at most $2^{64}$ keystream bits are to be generated. To the best of our understanding, this is sufficient for all
practical purposes.

\subsection{Effect of Shifts on Tap Positions \label{subsec-shift-tap-pos}}
In our concrete proposals, the number of variables $n=2m+1$ of the filtering function is about half the length $L$ of the LFSR (see Table~\ref{tab-Lm-pairs}). In fact, the
ratio $n/L$ in our proposals is much higher than similar ratios in all previous proposals. Using a high value of $n$ prevents certain well known attacks
as we discuss later. On the other hand, the high (compared to previous proposals) value of $n/L$ ratio means that the tap positions are placed close together.
Since $L-1$ bits of the next state of the LFSR are obtained by a shift of the previous
state, two states of the LFSR which are close in time share a number of bits. Further, with a high value of $n/L$ ratio, 
two keystream bits arising from two such nearby states may depend on a number of common state bits. This creates the possibility that in the XOR of two such
keystream bits, a number of terms involving the state bits cancel out. A bad choice of the tap positions can indeed lead to such an effect, as is explained
in Section~\ref{subsec-BV-attack}. Below we show that such cancellations do not occur in our proposals.

\paragraph{No cancellation of quadratic terms.}
From~\eqref{eqn-f}, the filtering function $f_{2m+1}(W,\mathbf{X},\mathbf{Y})$ has the term $\sym{Maj}_m(\mathbf{X})$. For $m\geq 6$, Theorem~\ref{thm-maj-prop}
assures us that the ANF of $\sym{Maj}_m(\mathbf{X})$ does not have any quadratic terms. All our concrete
proposals have $m$ (much) greater than 6, so we assume that $\sym{Maj}_m(\mathbf{X})$ does not contribute any quadratic term to the ANF of $f_{2m+1}$. 
Hence, the quadratic terms of $f_{2m+1}$ arise from the inner product $\langle\sym{rev}(\mathbf{X}),\mathbf{Y}\rangle$. 

When applied to the bits of the state, the inner product $\langle\sym{rev}(\mathbf{X}),\mathbf{Y}\rangle$ leads to $m$ quadratic terms
involving the state bits. Two states which are separated by $d$ time periods share a number of bits. More precisely, the rightmost $L-d$ bits
of the later state are the leftmost $L-d$ bits of the former state. This creates the possibility that 
some of the quadratic terms of the former state cancels with 
some of the quadratic terms of the later state. We show that due to our choice of the permutation $\pi$ as the bit reversal permutation $\sym{rev}$ and
the choice of placing the tap positions for $\mathbf{X}$ to the left of the tap positions for $\mathbf{Y}$, such cancellations do not occur. 

Let $\mathbf{s}^{(t)}$ be the state at time point $t$ as defined as in~\eqref{eqn-state-t}. The $p$-th quadratic term of $\mathbf{s}^{(t)}$ arising from
$\langle\sym{rev}(\mathbf{X}),\mathbf{Y})$ is $s_{L+t-1-i_p}s_{L+t-1-j_{m+1-p}}$, where $1\leq p\leq m$. Let
\begin{eqnarray}\label{eqn-qterm}
	\sym{qterm}(t,p) & = & \{L+t-1-i_p,L+t-1-j_{m+1-p}\},
\end{eqnarray}
i.e. $\sym{qterm}(t,p)$ is the 2-element set of indices which define the $p$-th quadratic term of $s^{(t)}$.

Let $t_1\geq 0$ and $d>0$ be integers, and $t_2=t_1+d$. Let $\mathbf{s}^{(t_1)}$ and $\mathbf{s}^{(t_2)}$ be the states at time points $t_1$ and $t_2$
respectively (obtained from~\eqref{eqn-state-t} by substituting $t_1$ and $t_2$ for $t$). If there are integers $p$ and $p^\prime$ with $1\leq p,p^\prime\leq m$ such 
that $\sym{qterm}(t_1,p)$ is equal to $\sym{qterm}(t_2,p^\prime)$, then the $p$-th quadratic term of $\mathbf{s}^{(t_1)}$
is identically equal to the $p^\prime$-th quadratic term of $\mathbf{s}^{(t_2)}$. In this case a cancellation of quadratic terms occurs between
$\mathbf{s}^{(t_1)}$ and $\mathbf{s}^{(t_2)}$. The following result shows that such cancellations do not occur.
\begin{proposition}\label{prop-no-cancel}
	For $0\leq t_1<t_2$ and $1\leq p,p^\prime\leq m$, $\sym{qterm}(t_1,p)\neq \sym{qterm}(t_2,p^\prime)$.
\end{proposition}
\begin{proof}
%
	Let $d=t_2-t_1>0$.
	We have $\sym{qterm}(t_1,p)=\{L+t_1-1-i_p,L+t_1-1-j_{m+1-p}\}$ and $\sym{qterm}(t_2,p^\prime)=\{L+t_2-1-i_{p^\prime},L+t_2-1-j_{m+1-p^{\prime}}\}$.
	The argument uses the condition on the integers $i_1,\ldots,i_m$ and $j_1,\ldots,j_m$ given in~\eqref{eqn-tap-pos}.
	Let if possible $\sym{qterm}(t_1,p)=\sym{qterm}(t_2,p^\prime)$. There are two cases to consider.

	\noindent{\em Case 1:} This case corresponds to the possibility 
	$L+t_1-1-i_p=L+t_2-1-i_{p^{\prime}}$ and $L+t_1-1-j_{m+1-p}=L+t_2-1-j_{m+1-p^{\prime}}$.
	The two conditions are respectively equivalent to $i_{p^{\prime}}=i_p+d$ and $j_{m+1-p^{\prime}}=j_{m+1-p}+d$. 
	Since $d>0$, from $i_{p^{\prime}}=i_p+d$ we have $i_{p^\prime}>i_p$ and from~\eqref{eqn-tap-pos}, we obtain $p^\prime>p$. 
	Similarly, from $j_{m+1-p^{\prime}}=j_{m+1-p}+d$, using $d>0$ 
	and~\eqref{eqn-tap-pos}, we obtain $m+1-p^{\prime}>m+1-p$, i.e. $p>p^\prime$, which is a contradiction to $p^\prime>p$.

	\noindent{\em Case 2:} This case corresponds to the possibility that $L+t_1-1-i_p=L+t_2-1-j_{m+1-p^{\prime}}$ and $L+t_2-1-i_{p^{\prime}}=L+t_1-1-j_{m+1-p}$.
	These two conditions are respectively equivalent to $j_{m+1-p^{\prime}}=i_p+d$ and $i_{p^{\prime}}=j_{m+1-p}+d$. Since $d>0$, from $i_{p^{\prime}}=j_{m+1-p}+d$
	we obtain $i_{p^{\prime}}>j_{m+1-p}$, but from~\eqref{eqn-tap-pos}, $i_{p^{\prime}}<j_{m+1-p}$, and so we obtain a contradiction.
\end{proof}


\paragraph{No cancellation of terms of the majority function.} 
As mentioned above, two calls to the majority function can have an overlap. This has the
potential of some terms in the ANF of the majority function arising from one of the calls cancelling out with terms in the ANF of the majority function
arising from the other calls. This is undesirable and reduces the ``complexity''. 
The next result provides a sufficient condition to prevent such undesirable cancellations.
\begin{proposition}\label{prop-maj-no-cancel}
	Let $m$ and $p$ be positive integers with $1\leq p<m$.
	Let $\mathbf{U}=(U_1,\ldots,U_p)$, $\mathbf{V}=(V_1,\ldots,V_{m-p})$ and $\mathbf{W}=(W_1,\ldots,W_{m-p})$. If $p\leq \floor{m/2}$, then
	no monomial in the ANF of $\sym{Maj}_m(\mathbf{U},\mathbf{V})$ cancels with any monomial in the ANF of $\sym{Maj}_m(\mathbf{U},\mathbf{W})$.
\end{proposition}
\begin{proof}
	From Theorem~\ref{thm-maj-prop}, we have that any monomial occuring in the ANF of $\sym{Maj}_m(\mathbf{U},\mathbf{V})$ has degree more than
	$\floor{m/2}$. So any monomial occurring in the ANF of $\sym{Maj}_m(\mathbf{U},\mathbf{V})$ must involve one of the $V_i$'s, and similarly,
	any monomial occurring in the ANF of $\sym{Maj}_m(\mathbf{U},\mathbf{W})$ must involve one of the $W_j$'s. Since the ANF of $\sym{Maj}_m(\mathbf{U},\mathbf{V})$
does not involve any $W_j$, and the ANF of $\sym{Maj}_m(\mathbf{U},\mathbf{W})$ does not involve any $V_i$, there can be no cancellation of terms between the
two ANFs.
\end{proof}
Proposition~\ref{prop-maj-no-cancel} assures us that if the overlap between two calls to majority is at most $\floor{m/2}$, then there is no cancellation of terms
arising from the ANF corresponding to the two calls. 

While defining $s^{(t)}$ in~\eqref{eqn-state-t}, we considered $s^{(0)}$ to be the state after the initialisation phase. More generally, by shifting the time steps,
we may consider $s^{(0)}$ to be the state at any point of time and $s^{(t)}$ to be the state obtained from $s^{(0)}$ after $t$ time steps. 
The function $f\circ\sym{proj}$ is applied to both the states $\mathbf{s}^{(0)}$ and $\mathbf{s}^{(t)}$. This in turn requires applying the majority function
to a subset of the bits of $\mathbf{s}^{(0)}$ and $\mathbf{s}^{(t)}$. More precisely, the functions calls
$\sym{Maj}(s_{L-1-i_1},\ldots,s_{L-1-i_m})$ and $\sym{Maj}(s_{L+t-1-i_1},\ldots,s_{L+t-1-i_m})$ occur corresponding to $\mathbf{s}^{(0)}$ and
$\mathbf{s}^{(t)}$ respectively. If $t<\kappa$, then there is a possibility that
the inputs to the two calls to majority have an overlap, i.e. some of the inputs are common to both the majority calls. Since majority is a symmetric function, the actual
positions where the common inputs occur do not matter. For $t\geq 1$, let
\begin{eqnarray}\label{eqn-nut}
	\nu_t & = & \#\left(\{L-1-i_1,\ldots,L-1-i_m\}\cap \{L+t-1-i_1,\ldots,L+t-1-i_m\} \right).
\end{eqnarray}
Note that $\nu_{t}=0$ for $t\geq \kappa$, and for $1\leq t<\kappa$,
\begin{eqnarray}\label{eqn-nut-posX}
	\nu_t & = & \sym{wt}(\sym{posX} \wedge (\sym{posX} \gg t)),
\end{eqnarray}
where $\wedge$ denotes the bitwise AND operation, and $\gg t$ denotes right shift by $t$ places.
Define
\begin{eqnarray}\label{eqn-nu}
	\nu & = & \max_{1\leq t <\kappa} \nu_t = \max_{1\leq t<\kappa} \sym{wt}(\sym{posX} \wedge (\sym{posX} \gg t)).
\end{eqnarray}
A design goal is to choose $\sym{posX}$ so as to minimise $\nu$. In particular, if $\nu$ is at most $\floor{m/2}$, then from Proposition~\ref{prop-maj-no-cancel}
there is no cancellation of terms arising from the two majority calls. Later we provide concrete values of $\nu$ (see Table~\ref{tab-nu-delta}) for our specific
proposals. For all cases, the condition $\nu\leq \floor{m/2}$ holds, and so the no cancellation property of terms arising from the ANF of the majority function
corresponding to the two calls is ensured.

\paragraph{Overlap of tap positions due to shifts.}
Let
\begin{eqnarray}\label{eqn-delta}
	\delta & = & \max_{1\leq t<2\kappa} \sym{wt}(\sym{pos} \wedge (\sym{pos} \gg t)).
\end{eqnarray}
Then $\delta$ is the maximum overlap between the tap positions that can be obtained by shifting the state by any value. 

\paragraph{Procedure for choosing the tap positions.} Suppose $\kappa$, $L\geq 2\kappa$, and $m$ are given. We need to determine $i_1,\ldots,i_m$
and $j_1,\ldots,j_m$, or equivalently, to determine $\sym{pos}$. (Recall that the tap position for $W$ is fixed to be $L-2\kappa$.) The 
leftmost $\kappa$ bits of $\sym{pos}$ is the string $\sym{posX}$. The value of $\nu$ is determined by $\sym{posX}$, while the value
of $\delta$ is determined by $\sym{pos}$. We would like to minimise $\nu$ as well as $\delta$. It is difficult to simultaneously minimise both.
From~\eqref{eqn-FSGA} (provided later), the value of $\delta$ is one of the factors relevant to protection against certain kinds of state guessing attacks. 
As explained in Remark~\ref{rem-FSGA} (also provided later), it is not essential to minimise $\delta$.
The parameter $\nu$, on the other hand, determines the size of the maximum overlap between two inputs to the majority function corresponding to two
different keystream bits. Minimising $\nu$ improves resistance to possible differential attacks (see Section~\ref{subsec-BV-attack}).
Since it is not essential to minimise $\delta$, we chose to focus on minimising $\nu$. 
The task of choosing $\sym{posX}$ such that $\nu$ is minimised is a combinatorial optimisation problem. There does not
seem to be any good way to obtain $\sym{posX}$ such that the corresponding $\nu$ is \textit{guaranteed} to be the minimum possible value. Instead, we adopted the
following procedure. Given $\kappa$, $L$ and $m$, we randomly generated the array $\sym{pos}$ in the following manner. Set $\sym{pos}[L-2\kappa]=1$ (the 
tap position for the variable $W$); set $i_1=0$, randomly select a subset $\{i_2,\ldots,i_m\}$ of size $m-1$ from $\{1,\ldots,\kappa-1\}$ and set
$\sym{pos}[L-1-i_p]=1$, for $p=1,\ldots,m$ (these provide the tap positions for the variables $X_1,\ldots,X_m$); set $j_m=2\kappa-2$, randomly select a
subset $\{j_1,\ldots,j_{m-1}\}$ from $\{\kappa,\ldots,2\kappa-3\}$ and set $\sym{pos}[L-1-j_{p^\prime}]=1$ (these provide the tap positions for 
the variables $Y_1,\ldots,Y_m$). From the obtained $\sym{pos}$, we extract $\sym{posX}$ and then we compute $\nu$ from $\sym{posX}$. So for each random
choice of $\sym{pos}$, we obtain a value of $\nu$. The random procedure is repeated 10000 times, providing 10000 values of $\nu$. 
Finally we return the $\sym{pos}$ for which the least value of $\nu$ is obtained. Then, for this $\sym{pos}$ compute the value of $\delta$. 
Through experimentation we found that repeating the random procedure more than 10000 times does not result in lower values of $\nu$. 
Later we provide the values of $\sym{pos}$, $\nu$ and $\delta$ for the concrete proposals that we make. 

Note that both $\nu$ and $\delta$ denote the maximum overlaps (of different kinds) that can arise due to shifts. The idea of tap position selection that we put forward is to 
minimise $\nu$ and $\delta$. While minimising both simultaneously is difficult, we call the general principle of minimising the maximum overlap due to shifts to be the 
\textit{shift overlap minimisation (SOM)} 
strategy of selecting tap positions. Selecting the tap positions to be a full positive difference set is a particular case of the SOM principle, where the minimum overlap is zero.
Since using tap positions which form a full positive difference set leads to the length of the shift register being too long, the SOM principle mandates that the maximum 
overlap due to shifts is as small as possible.

\subsection{Security Analysis \label{subsec-sec-anal}}
We consider the security of the stream cipher. For the analysis of security, we consider key recovery attacks, i.e. the goal of the adversary is to recover the
secret key. Since both the next state map $\sym{NS}$ (see~\eqref{eqn-next-st}) of the LFSR and the initialisation round function $\sym{IR}$ (see~\eqref{eqn-IR}) are invertible,
it follows that recovering the state of the stream cipher at any point of time essentially amounts to recovering the secret key, i.e. state recovery implies efficient key recovery. 

The parameters $L\geq 2\kappa$, $m$, and $\mu=\lfloor \sqrt{2\kappa}\rfloor$ are the design parameters, as are the tap positions encoded 
by $\sym{pos}$. The parameter $\nu$ is determined by $\sym{posX}$, while the parameter $\delta$ is determined by $\sym{pos}$. The primitive connection
polynomial $\tau(x)$ of the LFSR is also a parameter, as is the number of non-zero terms in $\tau(x)$. 
For a concrete choice targeted at a specific security level, the parameters need to be appropriately chosen so as to provide resistance against known classes of attacks.
Next we provide an overview of such attacks and determine the conditions on the parameters which provide resistance to the attacks.

We consider two basic \textit{attack parameters}, namely the number $N$ of keystream 
bits that is required for a successful attack, and the time complexity $T$ of the attack. By the time complexity we mean the number of bit operations required to
carry out the attack.
To ensure security at level $\kappa$ any attack should require $T>2^\kappa$. 
The condition $L\geq 2\kappa$ ensures that the size of the state is at least twice the size of the key. This prevents certain time/memory trade-off attacks.

Suppose $B$ is a parameter such that at most $2^B$ keystream bits are generated from a single key and IV pair. 
	So an attacker can access at most $2^B$ keystream bits. On the other hand, the attacker needs $N$ keystream bits to mount an attack. As a result, we have the 
	condition that $N$ is at most $2^B$, as otherwise the attack cannot be mounted. 
Further, a basic condition is that $N\leq T$, i.e. some operation is performed on each keystream bit that is required. So for $\kappa$-bit security using
$2^B$ keystream bits, we have $B\leq \kappa$. Recall that for our stream cipher proposal we have fixed $B=64$. For the description of the attacks we proceed without 
this restriction.

Below we perform a concrete analysis of known classes of attacks so as to be able to determine conditions on the parameters which ensure resistance to such attacks.
For the concrete analysis, we ignore small constants appearing in the expressions for $N$ and $T$ (i.e. we consider these constants to be 1) and focus only on the exponential 
components of these quantities.

\paragraph{Linear complexity attack.} The degree of $(\sym{Maj},\sym{rev})\mbox{-}\sym{MM}_{2m+1}$ is $d=2^{\lfloor \log_2 m\rfloor}$. If $L$ is a prime, then the 
linear complexity of the generated keystream sequence is ${L\choose d}$ (see~\cite{Ru86}) and so at least these many keystream bits are required to determine the linear complexity. 
So if 
\begin{eqnarray}\label{eqn-lin-com-resist}
	\alpha = {L\choose 2^{\lfloor \log_2 m\rfloor}} > 2^B,
\end{eqnarray}
then the linear complexity attack is not applicable.

\paragraph{Anderson leakage.}
An interesting method for exploiting leakage by the filtering function was introduced in~\cite{DBLP:conf/fse/Anderson94}. In this approach, the focus
is on knowing how much information is leaked by the filtering function \textit{even if the input LFSR sequence is replaced by a purely random bit sequence}.
Examples where such leakage can be observed were provided in~\cite{DBLP:conf/fse/Anderson94} for the case of filtering functions on a few variables. We call such
leakage to be Anderson leakage. Avoiding Anderson leakage amounts to showing that if the input sequence (to the filtering function) is purely random then the keystream 
sequence is also purely random. Theorem~2 in~\cite{DBLP:conf/fse/Golic96} provides a sufficient condition to ensure this property. The theorem states that if the $n$-variable 
filtering function is of the form $Z_1+g(Z_2,\ldots,Z_n)$, then there is no Anderson leakage. Since the filtering function $f$ (built from $\sym{MM}_{2m+1}$) that we 
propose is of the stated form, it does not exhibit Anderson leakage.

\paragraph{State guessing attacks.} 
These are guess-then-determine attacks. The idea is to guess some bits and then use the obtained keystream to verify the guess. The first attack
of this type was described in~\cite{DBLP:conf/fse/Golic96} and is called the inversion attack. The inversion attack is specifically applicable to filtering
functions of the form $W\oplus g(\mathbf{Z})$. Since the filtering function that we use is indeed of this form, we need to consider the resistance of the stream cipher
to the inversion attack. 

Suppose at some point of keystream generation, the state is $\mathbf{s}^{(0)}=(s_{L-1},\ldots,s_0)$ and the keystream bit generated from $\mathbf{s}^{(0)}$
is $w_0$. From the definition of the filtering function, the bit $w_0$ is generated from the bits $(s_{L-1},\ldots,s_{L-2\kappa})$, and we can write 
$w_0=s_{L-2\kappa}\oplus g(s_{L-1},\ldots,s_{L-2\kappa+1})$, where $g$ is some function (defined from $\sym{MM}_{2m}$ and is non-degenerate on $2m$ variables). 
Recall that the tap position for $X_1$ is $L-1$ and the tap position for $Y_m$ is $L-2\kappa+1$. 
The gap between the tap position of $X_1$ and the tap position of $Y_m$ (including both of these tap positions) is $\chi=2\kappa-1$. Note that
$s_{L-2\kappa}=w_0\oplus g(s_{L-1},\ldots,s_{L-2\kappa+1})$. Extending this equation in the backward direction, we obtain the following equations.
\begin{eqnarray*}
	s_{L-2\kappa} & = & w_0\oplus g(s_{L-1},\ldots,s_{L-2\kappa+1}) \\
	s_{L-2\kappa-1} & = & w_{-1}\oplus g(s_{L-2},\ldots,s_{L-2\kappa}) \\
	s_{L-2\kappa-2} & = & w_{-2}\oplus g(s_{L-3},\ldots,s_{L-2\kappa-1}) \\
	\cdots & \cdot & \cdots
\end{eqnarray*}
The sequence $w_0,w_{-1},w_{-2},\ldots$ is known. If we guess the values of $s_{L-1},\ldots,s_{L-2\kappa+1}$ (a total of $\chi$ bits), then we obtain $s_{L-2\kappa}$ from the first
equation, using the obtained value of $s_{L-2\kappa}$ in the second equation, we obtain the value of $s_{L-2\kappa-1}$, using the obtained value of $s_{L-2\kappa-1}$
in the third equation, we obtain the value of $s_{L-2\kappa-2}$, and so on. In other words, by guessing the values of $\chi$ of the state bits, we can obtain
values of $L-\chi$ previous state bits, giving us the values of all $L$ bits of the state at a point $L-\chi$ steps in the backward direction. Once the complete state is known,
the original guess can be verified by generating the keystream in the forward direction and matching with the obtained keystream. The method requires
$2^{\chi}$ guesses, and for each guess about $Lm$ bit operations are required. By our choice of the tap positions for $\mathbf{X}$ and $\mathbf{Y}$, the value
of $\chi$ is $2\kappa-1$. So the number of guesses required to successfully mount the attack is far greater than $2^{\kappa}$. This proves the resistance
of the stream cipher to the inversion attack. Note that the inversion attack is prevented due to $\chi$ being sufficiently large. This was one of the countermeasures to 
the inversion attack that was already proposed in~\cite{DBLP:conf/fse/Golic96}.

The inversion attack was later extended to the generalised inversion attack, the filter state guessing attack, and the generalised filter
state guessing attack~\cite{DBLP:conf/fse/Golic96,DBLP:journals/tc/GolicCD00,DBLP:journals/iet-ifs/HodzicPW19,DBLP:conf/balkancryptsec/PasalicHBW14,DBLP:journals/tit/WeiPH12}.
We briefly explain the idea behind this line of attacks.
Let $n=2m+1$ be the number of variables of the filtering function. At any point of time, each of the input bits to the filtering function can be written as a linear function
of the $L$ state bits of the LFSR obtained after the initialisation phase. So knowing the values of the $n$ input bits to the filtering function at any point of time provides
$n$ linear equations in $L$ variables. 
If the values of the input bits to the filtering function at $c$ points of time are known, where $nc\geq L$, then one obtains a system of $L$ linear equations in $L$ variables and 
hence can solve this system to obtain the initial state of the LFSR. Of course, one does not know the values of the input bits to the filtering function at any point of time. 
So a guessing strategy is used. 

Let as above the state of the LFSR at some point be $\mathbf{s}^{(0)}$ and the subsequent states of the LFSR be denoted as 
$\mathbf{s}^{(1)},\mathbf{s}^{(2)},\ldots,\mathbf{s}^{(c-1)}$. For $i=0,\ldots,c-1$, let $\mathbf{t}^{(i)}=\sym{proj}(\mathbf{s}^{(i)})$ and
$w_i=f_{2m+1}(\mathbf{t}^{(i)})$. Recall from the definition of $\delta$ that for $0\leq i<j\leq c-1$, $\mathbf{t}^{(i)}$ and $\mathbf{t}^{(j)}$ have an overlap 
of at most $\delta$ bits.
So in the $c$ $n$-bit strings $\mathbf{t}^{(0)},\ldots,\mathbf{t}^{(c-1)}$ there are at least $nc-\delta c(c-1)/2$ unknown bits and hence there are at least
$2^{nc-\delta c(c-1)/2}$ possibilities for these $c$ strings. Since $f_{2m+1}$ is balanced, knowledge of $w_0$ reduces the number of possibilities for
$\mathbf{t}^{(0)}$ by a factor of half, and more generally, the knowledge of $w_0,\ldots,w_{c-1}$ reduces the number of possibilities 
for $\mathbf{t}^{(0)},\ldots,\mathbf{t}^{(c-1)}$ by a factor of $2^c$. So at least $2^{(n-1)c-\delta c(c-1)/2}$ possibilities remain for the $c$ strings 
$\mathbf{t}^{(0)},\ldots,\mathbf{t}^{(c-1)}$.
For each of these possibilities, we obtain a system of $L$ equations in $L$ variables, the solution of which provides the value of initial state of the LFSR. 
The correctness of the obtained value of the state can be determined by generating the keystream from the obtained value and matching with the actual keystream.
Solving the system of $L$ linear equations requires $L^3$ bit operations, and generating the keystream from the obtained state requires about
$Ln$ bit operations. So the complexity of the attack is at least $2^{(n-1)c-\delta c(c-1)/2}(L^3+Ln)$ operations. Suppose the following condition holds.
\begin{eqnarray}\label{eqn-FSGA}
	2^{(n-1)c-\delta c(c-1)/2}(L^3+Ln) & > & 2^{\kappa} \mbox{ where $c$ is the least positive integer such that $nc\geq L$.}
\end{eqnarray}
Then the presently known state guessing attacks do not succeed at the $\kappa$-bit security level. 

\begin{remark}\label{rem-FSGA}
	In our concrete proposals given below in Table~\ref{tab-Lm-pairs}, the value of $c$ in~\eqref{eqn-FSGA} comes out to be 2, and so~\eqref{eqn-FSGA} becomes
	$2^{2(n-1)-\delta}(L^3+Ln)>2^{\kappa}$. Since the values of $n=2m+1$ in Table~\ref{tab-Lm-pairs} are quite close to $\kappa$, ensuring
	$2^{2(n-1)-\delta}(L^3+Ln)>2^{\kappa}$ does not require $\delta$ to be too small. In fact, the values of $\delta$ (and corresponding values
	of $L$, $m$ and $\kappa$) given in Table~\ref{tab-nu-delta} ensure that this inequality holds for all our proposals.
\end{remark}

Applying the attack requires knowing the pre-image sets of $0$ and $1$ of the function $f_{2m+1}$. These need to be stored separately, and depending upon the value
of $w_i$, the appropriate pre-image set is to be used. So the storage required is $2^{2m+1}$. For the concrete proposals that we put forward later, the value of
$n=2m+1$ is quite close to that of the security level $\kappa$ (see Table~\ref{tab-Lm-pairs}). As a result, the memory requirement will be prohitively high to apply the attack.

The complexity of the above attack depends on the value of $\delta$, which is the maximum overlap of tap positions between shifts of the state vector.
The recommendations for tap positions
in the literature~\cite{DBLP:conf/fse/Golic96,DBLP:journals/tc/GolicCD00,DBLP:journals/iet-ifs/HodzicPW19,DBLP:conf/balkancryptsec/PasalicHBW14,DBLP:journals/tit/WeiPH12}
aim to reduce this overlap. A commonly used recommendation is full positive difference set, i.e. the absolute values of the differences between the tap positions
should be distinct. For a filtering function with $n$ variables, this recommendation results in the gap between the first and the last tap positions
to be more than $1+2+\cdots+(n-1)=n(n-1)/2$ (since the successive differences must be distinct). So the size of the LFSR is at least quadratic in $n$. From efficiency
considerations, this forces the value of $n$ to be small. By not following this recommendation, we have done away with the condition that $L$ must be at least
quadratic in $n$. Instead, by considering the fundamental requirement behind state guessing attacks, we identified~\eqref{eqn-FSGA} as the condition to resist
such attacks. This allows us to choose $n$ to be quite large and close to $\kappa$. Of course, this is possible due to the fact that we have an efficient
method for implementing the filtering functions on such large values of $n$.

\paragraph{Fast correlation attacks.}
The basic correlation attack~\cite{DBLP:journals/tc/Siegenthaler85} is applicable to the combiner generator model. Applying this attack to the filter generator model results in 
going through all the possible $2^L$ states of the LFSR. Since by our choice $L\geq 2\kappa$, the basic correlation attack does not defeat the $\kappa$-bit 
security level. (An early correlation attack~\cite{DBLP:conf/eurocrypt/Siegenthaler85} on the nonlinear filter generator finds an equivalent representation
of the stream cipher when the filtering function is not known; since we assume that the filtering function is known, this attack is not relevant to our context.)

Fast correlation attacks do not require exhaustive search on the states of the LFSR. There is a large literature on fast correlation attacks including
older papers such as~\cite{DBLP:journals/joc/MeierS89,DBLP:conf/crypto/JohanssonJ99,DBLP:conf/eurocrypt/JohanssonJ99,DBLP:conf/fse/ChepyzhovJS00,DBLP:conf/eurocrypt/CanteautT00,DBLP:journals/ipl/JonssonJ02,CF02,DBLP:conf/eurocrypt/ChoseJM02} as well as more recent papers such 
as~\cite{DBLP:conf/crypto/TodoIMAZ18,DBLP:journals/tosc/ZhouFZ22,DBLP:journals/tosc/ZhangLGJ23,DBLP:journals/tit/MaJGCS24,DBLP:journals/dcc/MartinezS24}. 
See~\cite{Can05,DBLP:reference/crypt/Canteaut11d,DBLP:conf/fse/Meier11,DBLP:journals/ccds/AgrenLHJ12} for surveys of the area. 
In the following, we evaluate security against some representative fast correlation attacks and show how to choose the values of the design parameters so that these attacks
can be resisted.
Recall from Proposition~\ref{prop-MM-odd-nl-deg} that for $\sym{MM}_{2m+1}$, and hence for $f_{2m+1}$, the linear bias $\varepsilon=2^{-m-1}$. \\


\noindent{\textit{Type of attack-I}.} For attacks based on low-weight parity-check equations~\cite{DBLP:journals/joc/MeierS89,DBLP:conf/eurocrypt/CanteautT00}, the 
number of keystream bits required is about $N=(2\varepsilon)^{-2(d-2)/(d-1)}\cdot 2^{L/(d-1)}$, the pre-computation step requires about $N^{d-2}/(d-2)!$ operations, 
and the (online) time for decoding is about $(2\varepsilon)^{-2d(d-2)/(d-1)}\cdot 2^{L/(d-1)}$, where $d\geq 3$ is the number of non-zero terms in (some multiple of the) 
LFSR connection polynomial, and $\varepsilon$ is the linear bias. Note that for this attack, the parameter $d$ is not the algebraic degree of the filtering function.

\noindent{\textit{Evaluation against $\mathcal{S}(L,m)$}.} We have $\varepsilon=2^{-m-1}$ and so $N=2^{2m(d-2)+L)/(d-1)}$.
We set $N=2^B$ and so $(d-1)(B-2m)=L-2m$. If $B=2m$, then $L=2m$. We choose $L$ and $m$ to ensure that $L>2m$, and so $B\neq 2m$. In this
case, we solve for $d$ to obtain $d=1+(L-2m)/(B-2m)$. If $B<2m$, then $d<1$ which violates the condition $d\geq 3$ and the attack does not work. 
So let us consider $B>2m$. In this case, substituting the obtained expression for $d$ in the expression for the decoding time, we find the decoding time to be 
$2^{(2mL+B(B-4m))/(B-2m)}$. So the following condition ensures $\kappa$-bit security.
\begin{eqnarray}\label{eqn-kappa-low-wt-FCA}
	L>2m \mbox{ and either } B<2m \mbox{ or } \kappa < \frac{2mL+B(B-4m)}{B-2m}. 
\end{eqnarray}
Using the relation between $d$ and $B$, we have $(2mL+B(B-4m))/(B-2m) = 2m(d-1) + (L-2m)/(d-1)>4m$ for $d\geq 3$. So if $\kappa<4m$, 
then $\kappa$-bit security is achieved \textit{for all values of $d\geq 3$}. In particular, it does not matter whether the
feedback polynomial of the LFSR is sparse, or whether it has a sparse multiple. In all our concrete proposals, the condition $\kappa<4m$ holds (see Table~\ref{tab-Lm-pairs}). \ \\

\noindent{\textit{Type of attack-II}.}
For attacks based on general decoding,~\cite{Can05} identifies the key idea to be from~\cite{DBLP:conf/fse/ChepyzhovJS00}. 
The attack in~\cite{DBLP:conf/fse/ChepyzhovJS00} requires $N$ to be
about $\varepsilon^{-2}\cdot 2^{(L-k)/w}$ under the condition $N \gg (2\varepsilon)^{-2w}$, and the time complexity of the decoding step is about
$2^k\cdot (2\varepsilon)^{-2w}$, where $\varepsilon$ is the linear bias of the filtering function, and $k\in\{1,\ldots,L\}$ and $w\geq 2$ are algorithm parameters.

\noindent{\textit{Evaluation against $\mathcal{S}(L,m)$}.} 
Again we have $\varepsilon=2^{-m-1}$. Setting $N=2^B$, we obtain $k=L+2w(m+1)-wB$. Setting $T$ to be the time complexity of the
decoding step, we obtain $\log_2T=L+2w(2m+3)-wB$. Since $L\geq 2\kappa$, ensuring $2(2m+3)\geq B$ is (more than) sufficient to ensure $\kappa$-bit security,
\textit{irrespective of the values of $w$ and $k$}. We record this condition as follows.
\begin{eqnarray}\label{eqn-kappa-gen-decode-FCA}
	B & \leq & 2(2m+3).
\end{eqnarray}
\\
\noindent{\textit{Type of attack-III}.} The attacks in~\cite{DBLP:journals/ipl/JonssonJ02,CF02} apply specifically to the filter generator model. These two attacks
are essentially the same when the filtering function is plateaued (which is the case for $f_{2m+1}$). For the attack, $N$ is about $2^{(L-k)/w}$, and 
$T$ is about $2^k\cdot F^w$, where $w$ and $k$ are as in the
attack in~\cite{DBLP:conf/fse/ChepyzhovJS00} (see above) and $F$ is the size of the support of the Walsh transform of the filtering function.

\noindent{\textit{Evaluation against $\mathcal{S}(L,m)$}.} 
The size of the support of the Walsh transform of $\sym{MM}_{2m+1}$ is equal to the size of the support of the Walsh transform of $\sym{MM}_{2m}$. 
Since $\sym{MM}_{2m}$ is bent, it follows that the size of the support of $\sym{MM}_{2m}$ is $2^{2m}$. So $F=2^{2m}$. Setting $N=2^B$, we obtain 
$k=L-wB$. Substituting $k$ in the expression for $T$, we obtain $\log_2T=L+w(2m-B)$. Since $L\geq 2\kappa$ and $w\geq 2$, we have
$\log_2T=L+w(2m-B)\geq 2\kappa + 2(2m-B)$. So if $2\kappa + 2(2m-B) > \kappa$, or equivalently, $\kappa+4m>2B$, then $\kappa$-bit security is achieved against the attack.
We record this condition as follows.
\begin{eqnarray}\label{eqn-kappa-gen-CF-FCA}
	B & < & (\kappa+4m)/2.
\end{eqnarray}
\\
\noindent \textit{Other fast correlation attacks.} We next consider some of the more recent attacks.
The attack in~\cite{DBLP:conf/crypto/TodoIMAZ18} is based on using $M>1$ linear approximations. For this attack, both $N$ and $T$ are about $2^{L-\beta}$,
where $\beta$ is an algorithm parameter. A necessary condition for the attack to succeed is that $M>2^\beta$. 
For the attack to succeed at the $\kappa$-bit security level, i.e. $T\leq 2^\kappa$, it is required to have $\beta\geq L-\kappa$. From the bound on $M$, it follows 
that more than $2^{L-\kappa}$ linear approximations with sufficiently high correlations are required. 
For Grain-128a~\cite{DBLP:journals/ijwmc/AgrenHJM11}, $L$ equals 128 and about $2^{26.58}$ (i.e. $M$ is about $2^{26.58}$) linear approximations 
with absolute correlations at least $2^{-54.2381}$ were identified in~\cite{DBLP:conf/crypto/TodoIMAZ18}.
For $\kappa=128$, Table~\ref{tab-Lm-pairs} recommends $L=257$ and $m=58$. 
To apply the attack in~\cite{DBLP:conf/crypto/TodoIMAZ18} to $\mathcal{S}(257,59)$, more than $2^{129}$ linear approximations with sufficiently high correlations are required. 
The linear bias of the filtering function itself is $2^{-60}$. Finding multiple linear approximations requires combining keystream bits which further lowers the linear bias. 
So there is no approximation with linear bias greater than $2^{-60}$. 
Further, going through the details of the attack in~\cite{DBLP:journals/ijwmc/AgrenHJM11}, we could not identify any method to obtain more than $2^{129}$ linear approximations.
So there does not seem to be any way to apply the attack in~\cite{DBLP:conf/crypto/TodoIMAZ18} to $\mathcal{S}(257,59)$.

Subsequent works (such as~\cite{DBLP:journals/tosc/ZhouFZ22,DBLP:journals/tosc/ZhangLGJ23,DBLP:journals/tit/MaJGCS24,DBLP:journals/dcc/MartinezS24})
on fast correlation attacks use vectorial decoding technique along with multiple linear approximations, use of the BKW algorithm~\cite{DBLP:journals/jacm/BlumKW03},
and multivariate correlation attack. The attacks are quite complex and there are no simple closed form expressions for the values of $N$ and $T$. 
The stream ciphers to which these attacks are applied are Grain-128a and Sosemanuk~\cite{sosemanuk}. 
From the above discussion, we already know that the linear bias of $\mathcal{S}(257,59)$ is substantially lower than that of Grain-128a. For 
Sosemanuk, the best known~\cite{DBLP:journals/tit/MaJGCS24} linear approximation has correlation $2^{-20.84}$, which is far greater than the linear bias
$2^{-60}$ for $\mathcal{S}(257,59)$. The very low linear bias of $\mathcal{S}(257,59)$ and in general of the other stream cipher proposals in Table~\ref{tab-Lm-pairs}
make the attacks in~\cite{DBLP:journals/tosc/ZhouFZ22,DBLP:journals/tosc/ZhangLGJ23,DBLP:journals/tit/MaJGCS24,DBLP:journals/dcc/MartinezS24} inapplicable
at the stated security levels.


\paragraph{(Fast) Algebraic attacks.} For a filtering function $f$ having $\sym{AI}(f)=a$, an algebraic attack requires about
$N=\sum_{i=0}^a{L\choose i}$ keystream bits and has time complexity $T$ to be about $N^\omega$, where $\omega$ is the exponent of matrix multiplication 
(see Section~3.1.5 of~\cite{BF-book}). Following~\cite{DBLP:conf/eurocrypt/CourtoisM03} we take the value of $\omega$ to be given by Strassen's algorithm, 
i.e. $\omega=\log_27\approx 2.8$. 
For our proposal, the filtering function is $f_{2m+1}$ whose algebraic resistance is the same as that of $(\sym{Maj},\sym{rev})\mbox{-}\sym{MM}_{2m+1}$. 
Let $a=\sym{AI}(f_{2m+1})=\sym{AI}((\sym{Maj},\sym{rev})\mbox{-}\sym{MM}_{2m+1})$. From Proposition~\ref{prop-AI-n}, we assume that $a=\lceil m/2\rceil$, i.e.
the actual algebraic immunity of $f_{2m+1}$ is equal to the lower bound on the algebraic immunity. Let 
\begin{eqnarray}\label{eqn-AA-resist}
	\beta & = & \left(\sum_{i=0}^{\lceil m/2\rceil}{L\choose i}\right)^{2.8}.
\end{eqnarray}
So choosing $L$ and $m$ such that $T=\beta> 2^{\kappa}$ prevents algebraic attacks at the $\kappa$-bit security level.

For any $(e,d)$ for which there are non-zero functions $g$ and $h$ of degrees $e$ and $d$ respectively such that $gf_{2m+1}=h$, 
the Berlekamp-Massey step in a fast algebraic attack on the filter generator model has time 
complexity $\mathcal{O}(ED\log D)$, where $E=\sum_{i=0}^e{L\choose i}$ and $D=\sum_{i=0}^d{L\choose i}$ (see~\cite{DBLP:conf/crypto/HawkesR04} and Section~3.1.5 
of~\cite{BF-book}). This complexity dominates the overall time complexity of a fast algebraic attack. Ignoring the logarithm term, we take $T$ to be equal to $ED$. 
The number $N$ of keystream bits required is about $2E$. Since $h$ is non-zero, the maximum value of $e$ is one less than the algebraic immunity $a$. 
From Proposition~\ref{prop-AI-n}, we again assume as above that $a=\lceil m/2\rceil$. Recall that for any Boolean function, its fast algebraic immunity is at least one 
more than its algebraic immunity, i.e. $\sym{FAI}(f_{2m+1})\geq a+1$. For any functions $g$ and $h$ of degrees $e$ and $d$ respectively such that 
$gf_{2m+1}=h$, we have $e+d\geq \sym{FAI}(f_{2m+1})\geq a+1$, i.e. $d\geq a+1-e$. 
Define
\begin{eqnarray}\label{eqn-FAA-resist}
	\gamma & = & \min_{\stackrel{1\leq e\leq a-1,}{d=a+1-e}}\left(\sum_{i=0}^e{L\choose i}\right) \left(\sum_{i=0}^d{L\choose i} \right).
\end{eqnarray}
So choosing $L$ and $m$ such that $T=\gamma> 2^{\kappa}$ prevents fast algebraic attacks at the $\kappa$-bit security level.

\subsection{Concrete Choices of $L$, $m$ and $\sym{pos}$ \label{subsec-concrete-ch}}
Given $\kappa$, we obtained representative values of $L$ and $m$. The procedure we followed to obtain $L$ and $m$ is to choose the value of $L$ to be the first prime number 
greater than $2\kappa$, and then for the chosen value of $L$, choose $m$ to be the least integer such that both $\beta$ and $\gamma$ (given 
by~\eqref{eqn-AA-resist} and~\eqref{eqn-FAA-resist} respectively) are greater than $2^\kappa$. This ensures security against (fast) 
algebraic attacks considered above. 
Next, using $\kappa$ and the corresponding values of $L$ and $m$, we computed the maximum value of $B$ such that $B\leq \kappa$, $\alpha >2^B$,
and~\eqref{eqn-kappa-low-wt-FCA},~\eqref{eqn-kappa-gen-decode-FCA} and~\eqref{eqn-kappa-gen-CF-FCA} hold. In each of the cases that we considered, it
turns out that this maximum value of $B$ is in fact $\kappa$. So for the chosen values of $L$ and $m$, the corresponding $\mathcal{S}(L,m)$ ensures $\kappa$-bit security 
against the above analysed correlation attacks even when the adversary has access to $2^{\kappa}$ keystream bits. Note that we allow at most $2^{64}$ keystream bits
to be generated from a single key and IV pair. Since $64<\kappa$, it follows that the concrete proposals are indeed secure against that fast correlation attacks
considered above.
Table~\ref{tab-Lm-pairs} shows the values of $\kappa$, $L$, $m$ and other parameters of the filtering function.
We note the following points regarding the entries in Table~\ref{tab-Lm-pairs}.


\begin{table}
\centering
	{\scriptsize
	\begin{tabular}{|c||c|c||c|c|c|c|c|}
		\hline
		$\kappa$ & $L$ & $m$ & $2m+1$ & deg & $\sym{LB}$ & $\sym{AI}$ & $\sym{FAI}$ \\ \hline
		 80      & 163 & 37 & 75 & 32 & $2^{-38}$ & 19 & 20 \\ \hline
		128      & 257 & 59 & 119 & 32 & $2^{-60}$ & 30 & 31 \\ \hline 
		160      & 331 & 71 & 143 & 64 & $2^{-72}$ & 36 & 37 \\ \hline
		192      & 389 & 87 & 175 & 64 & $2^{-88}$ & 44 & 45 \\ \hline
		224      & 449 & 101 & 203 & 64 & $2^{-102}$ & 51 & 52 \\ \hline
		256      & 521 & 115 & 231 & 64 & $2^{-116}$ & 58 & 59 \\ \hline 
	\end{tabular}
	\caption{Values of $L$ and $m$ which provide $\kappa$-bit security against the attacks analysed in this section when the attacker has
	access to at most $2^B$ keystream bits. \label{tab-Lm-pairs} }
	}
\end{table}
\begin{compactenum}
\item The number of variables of the filtering function $f_{2m+1}$ is $2m+1$ (and not $m$). 
\item For each $\kappa$, the value of $L$ is the first prime number greater than $2\kappa$. Our choice of a prime number for the value of $L$ is to ensure that the 
	linear complexity of the generated keystream is indeed equal to $\alpha$.
\item The table provides representative values of $L$ and $m$. It is possible to have other pairs of values for $(L,m)$ which provide $\kappa$-bit security. 
	Different values of $L$ and $m$ lead to different sizes of the circuits implementing the corresponding stream ciphers. For a fixed value of $\kappa$,
	a practical designer may need to consider various values of $(L,m)$ to determine which pair provides the smallest circuit. In Section~\ref{sec-MM-eff},
	we provide gate count estimates for the stream ciphers corresponding to the values of $L$ and $m$ in the table.
\item For each entry in the table, our calculation shows that $\beta\gg \gamma$. For example, for $\mathcal{S}(257,59)$, we obtained
	$\beta=2^{364.38}$ while $\gamma=2^{130.12}$. This is not surprising, since the fast algebraic attack is known to be much faster than the
	basic algebraic attack.
\item The feedback polynomial for the LFSR has to be a primitive polynomial of degree $L$. The above analysis of correlation attacks shows that for the obtained values
	of $L$ and $m$, it does not matter whether the polynomial is sparse, or whether it has a sparse multiple. So from an efficiency point of view, one may 
	choose a low weight primitive polynomial. Concrete choices of primitive polynomials for the values of $L$ in Table~\ref{tab-Lm-pairs} are shown 
	in Table~\ref{tab-prim-poly}.
\end{compactenum}

\begin{table}
\centering
	{\scriptsize
	\begin{tabular}{|l|l|}
		\hline 
		$L$ & prim poly \\ \hline
		163 &	$x^{163} \oplus x^7 \oplus x^6 \oplus x^3 \oplus 1$ \\ \hline
		257 & 	$x^{257} \oplus x^7 \oplus x^5 \oplus x^4 \oplus x^3 \oplus x^2 \oplus 1$ \\ \hline
		331 & $x^{331} \oplus x^7 \oplus x^6 \oplus x^5 \oplus x^4 \oplus x^2 \oplus 1$ \\ \hline
		389 & $x^{389} \oplus x^7 \oplus x^6 \oplus x^3 \oplus x^2 \oplus x \oplus 1$ \\ \hline
		449 & $x^{449} \oplus x^9 \oplus x^6 \oplus x^5 \oplus x^4 \oplus x^3 \oplus x^2 \oplus x \oplus 1$ \\ \hline
		521 & $x^{521} \oplus x^9 \oplus x^6 \oplus x^5 \oplus x^3 \oplus x \oplus 1$ \\ \hline
	\end{tabular}
	\caption{Concrete choices of primitive polynomials for the values of $L$ shown in Table~\ref{tab-Lm-pairs}. \label{tab-prim-poly} }
	}
\end{table}

\paragraph{Concrete choices of the tap positions.}
Given the values of $L$ and $m$ (and also $\kappa$), we applied the procedure for choosing the tap positions encoded by $\sym{pos}$ as
outlined in Section~\ref{subsec-shift-tap-pos}. The leftmost $\kappa$ bits of $\sym{pos}$ is the string $\sym{posX}$. The next $\kappa-1$ bits of $\sym{pos}$
(i.e. the positions $\sym{pos}[L-\kappa-1,\ldots,L-2\kappa+1]$) encode the tap positions for the variables in $\mathbf{Y}$. We define
$\sym{posY}$ to be the $\kappa$-bit string formed by appending a 0 to the $(\kappa-1)$-bit segment $\sym{pos}[L-\kappa-1,\ldots,L-2\kappa+1]$ of $\sym{pos}$.
So both $\sym{posX}$ and $\sym{posY}$ are $\kappa$-bit strings and are uniquely defined from $\sym{pos}$. Further, given $\sym{posX}$ and $\sym{posY}$
it is possible to uniquely construct $\sym{pos}$ (by concatenating $\sym{posX}$ and $\sym{posY}$, appending $L-2\kappa$ zeros, and setting the bit in 
position $L-2\kappa$ to be 1, to encode the tap position for $W$). So providing $\sym{posX}$ and $\sym{posY}$ is equivalent to providing $\sym{pos}$.
In Table~\ref{tab-posX-posY} we provide the values of $\sym{posX}$ and $\sym{posY}$ encoded as hexadecimal strings. 
For example, the entry for $\sym{posX}$ corresponding to $\mathcal{S}(257,59)$ is {\tt be352}$\cdots$ which encodes the binary string
10111110001101010010$\cdots$.
In Table~\ref{tab-nu-delta} we provide the corresponding values of $\nu$ and $\delta$. Recall that $\nu$ is obtained from $\sym{posX}$
as in~\eqref{eqn-nu} and $\delta$ is obtained from $\sym{pos}$ as in~\eqref{eqn-delta}.

The number of variables of the filtering function is $n=2m+1$. 
For each of the obtained values of $L$ and $n$, and the corresponding value of $\delta$ we checked whether~\eqref{eqn-FSGA} holds.
In each case we found that~\eqref{eqn-FSGA} indeed holds. So the stream cipher proposals ensure security against state guessing attacks.

\begin{table}
\centering
	{\scriptsize 
	\begin{tabular}{|l|l|}
		\hline
		$\mathcal{S}(L,m)$ & \multicolumn{1}{|c|}{$\sym{posX}$ and $\sym{posY}$} \\ \hline
		\multirow{2}{*}{$\mathcal{S}(161,37)$} 
		& {\tt d569a664f500763506c3} \\
		& {\tt ff0149d4c640e9846cf2} \\ \hline
		\multirow{2}{*}{$\mathcal{S}(257,59)$} 
		& {\tt be352d9ca349432b80b38ac54e5164c9} \\
		& {\tt d2ece08cbb5566d608a69b19e4a91418} \\ \hline
		\multirow{2}{*}{$\mathcal{S}(331,71)$} 
		& {\tt ea4308e1229305d185450cfa26b0dcac68c4ab7d} \\
		& {\tt 1dbb5a438e7e55904cc04406bf0670ad728462b0} \\ \hline
		\multirow{2}{*}{$\mathcal{S}(389,87)$} 
		& {\tt a0265ea181b73a460fb50d8482590e584d15869de343957e} \\ 
		& {\tt c6b218be600d6183c074d00fde24e1c308ebb06cebab0f84} \\ \hline
		\multirow{2}{*}{$\mathcal{S}(449,101)$} 
		& {\tt e9507d49d4f4609a710d8d291eb466430af5668b03ec424c18417d86} \\
		& {\tt d288451f8f0554a46615f4448afa34aab8673d0647044afcd4682ec4} \\ \hline
		\multirow{2}{*}{$\mathcal{S}(521,115)$} 
		& {\tt c1ec835120741f290154b122618c625f0a9e77c5172cac84ae564390b2e91fda} \\
		& {\tt 5865fda7830eca37d0c2045994e9c83b1c55e13f1966c220809bc019d37f0054} \\ \hline
	\end{tabular}
	\caption{The strings $\sym{posX}$ and $\sym{posY}$ corresponding to the values of $L$ and $m$ for $\kappa$ which are shown in Table~\ref{tab-Lm-pairs}. 
	The first row corresponds to $\sym{posX}$ and the second row corresponds to $\sym{posY}$. \label{tab-posX-posY} }
	}
\end{table}

\begin{table}
\centering
	{\scriptsize
	\begin{tabular}{|c|c|c|}
		\hline
		$\mathcal{S}(L,m)$ & $\nu$ & $\delta$ \\ \hline
		$\mathcal{S}(161,37)$ & 16 & 36 \\ \hline
		$\mathcal{S}(257,59)$ & 26 & 57 \\ \hline
		$\mathcal{S}(331,71)$ & 30 & 69 \\ \hline
		$\mathcal{S}(389,87)$ & 39 & 86 \\ \hline
		$\mathcal{S}(449,101)$ & 45 & 96 \\ \hline
		$\mathcal{S}(521,115)$ & 51 & 112 \\ \hline
	\end{tabular}
	\caption{Values of $\nu$ and $\delta$ corresponding to the strings $\sym{posX}$ and $\sym{pos}$ (built from $\sym{posX}$ and
	$\sym{posY}$) shown in Table~\ref{tab-posX-posY}. \label{tab-nu-delta} }
	}
\end{table}

Given $\sym{pos}$ and the coefficient vector $\mathbf{c}$ of the primitive connection polynomial, it is required to determine the string $\mathbf{d}$ which provides
the feedback positions for the initialisation round function $\sym{IR}$ (see~\eqref{eqn-IR}). 
The string $\mathbf{d}$ has to be chosen so as to satisfy the conditions given by~\eqref{eqn-feed-back-cond}.
The weight of $\mathbf{d}$ is equal to $\mu=\lfloor\sqrt{2\kappa}\rfloor$. In Table~\ref{tab-d-concrete} we provide concrete choices of $\mathbf{d}$ by listing the positions
where $\mathbf{d}$ is 1.
\begin{table}
\centering
	{\scriptsize 
	\begin{tabular}{|l|l|}
		\hline
		$\mathcal{S}(L,m)$ & \multicolumn{1}{|c|}{$\mathbf{d}$} \\ \hline
		$\mathcal{S}(163,37)$ & 30 42 55 67 84 90 104 114 130 138 150 162 \\ \hline
		$\mathcal{S}(257,59)$ & 17 33 48 64 81 96 112 129 145 161 178 194 209 224 241 256 \\ \hline
		$\mathcal{S}(331,71)$ & 25 43 61 78 98 115 134 151 168 186 206 224 242 259 277 294 312 330 \\ \hline
		$\mathcal{S}(389,87)$ & 46 65 84 104 123 141 160 179 203 217 236 256 274 293 314 331 350 369 388 \\ \hline
		$\mathcal{S}(449,101)$ & 29 49 70 94 112 133 155 175 197 218 239 260 282 303 322 344 364 386 408 431 448 \\ \hline
		$\mathcal{S}(521,115)$ & 58 80 102 125 146 169 192 212 235 256 278 301 324 344 371 388 412 433 454 479 498 520 \\ \hline
	\end{tabular}
	\caption{Concrete choices of $\mathbf{d}$ corresponding to the various choices of $\sym{pos}$ obtained from $\sym{posX}$ and $\sym{posY}$ in Table~\ref{tab-posX-posY},
	and the choices of the coefficient vector $\mathbf{c}$ in Table~\ref{tab-prim-poly}. \label{tab-d-concrete}}
	}
\end{table}

\subsection{Differential Attacks \label{subsec-BV-attack}}
Choosing the filtering function to protect against certain known classes of attacks does not, however, protect against all possible attacks.
There are attacks which can succeed even if the filtering function is properly chosen. One such attack on a previous version of the proposal
was described in~\cite{cryptoeprint:2025/197}. To describe the attack and the design modification to resist it we need to mention the aspects of the
design which were different in the previous proposal. There are three such aspects.
\begin{enumerate}
	\item In the definition of $\sym{MM}_{2m}$ in the earlier version we had chosen $\pi$ to be the identity permutation, whereas in the
		present version we choose $\pi$ to be the bit reversal permutation.
	\item In the previous version, the tap positions were chosen as follows. From the state $(s_{L-1},\ldots,s_0)$ of the LFSR, the value of $W$ in 
		$f_{2m+1}(W,\mathbf{X},\mathbf{Y})$ was chosen to be the state bit $s_{L-\kappa+m}$, the values of the variables in $\mathbf{X}$ were chosen to be the 
		state bits $s_{L-\kappa+m-1},\ldots,s_{L-\kappa}$, and the values of the variables in $\mathbf{Y}$ were chosen to be the state bits 
		$s_{L-2\kappa+m-1},\ldots,s_{L-2\kappa}$. 
	\item In the previous version, during the initialisation phase, the feedback from the filtering function was fed back into the state by
		XORing it with the next bit of the LFSR, i.e. it was fed back to only one position of the state.
\end{enumerate}
Suppose the state after the initialisation phase is $(s_{L-1},\ldots,s_0)$, and for $t\geq 1$, let $s_t=\sym{nb}(s_{t-1},\ldots,s_{t-L})$. Let the keystream bit
generated from the state $(s_t,s_{t-1},\ldots,s_{t-L+1})$ be $w_t$. Using the definition of $\pi$ to be the identity permutation and the tap positions as mentioned
above, we have 
\begin{samepage}
\begin{eqnarray*}
	w_t & = & s_{t-\kappa+m}\oplus \langle 
	\pi(s_{t-\kappa+m-1},\ldots,s_{t-\kappa+1},s_{t-\kappa}),
	(s_{t-2\kappa+m-1},\ldots,s_{t-2\kappa+1},s_{t-2\kappa})\rangle \\
	& & \quad 
	\oplus \sym{Maj}_m(s_{t-\kappa+m-1},\ldots,s_{t-\kappa+1},s_{t-\kappa}), 
\end{eqnarray*}
\end{samepage}
So 
\begin{eqnarray*}
	w_t \oplus w_{t+1}
	& = & s_{t-\kappa+m-1}s_{t-2\kappa+m-1} \oplus s_{t+1-\kappa}s_{t+1-2\kappa} \\
	& & \oplus \sym{Maj}_m(s_{t-\kappa+m-1},\ldots,s_{t-\kappa+1},s_{t-\kappa}) \oplus \sym{Maj}_m(s_{t-\kappa+m},\ldots,s_{t-\kappa+2},s_{t-\kappa+1}).
\end{eqnarray*}
In other words, due to the choice of $\pi$ as the identity permutation, the selection of tap positions as consecutive positions of the state, 
and the fact that two successive keystream bits are obtained from a single shift of the LFSR sequence,
$m-1$ of the quadratic terms in the inner product cancelled out when $w_t$ and $w_{t+1}$ are XORed together. Further, the inputs to the two calls to 
$\sym{Maj}$ have an overlap of size $m-1$. These features were observed in~\cite{cryptoeprint:2025/197}
and exploited to mount a differential attack. The idea of the differential attack is to introduce a difference in the top most bit position of the IV. 
During the initialisation phase, this difference travels unchanged for $\kappa-m$ steps and then produces a difference in the input to the filtering function.
The difference in the output of the filtering function is fed back into the leftmost bit of the LFSR. As the LFSR is further shifted, this difference then travels without 
any further modification creating a high probability truncated differential. Combined with the simplified form of $w_t\oplus w_{t+1}$, this leads to an efficient
key recovery attack.

In the present version, we have chosen $\pi$ to be the bit reversal permutation, and all the tap positions for $\mathbf{X}$ to occur to the left of all the tap
positions for $\mathbf{Y}$. As a result, in the XOR of any number of keystream bits, no quadratic term cancels out (see Proposition~\ref{prop-no-cancel}). So, in particular, 
the above kind of cancellation of quadratic terms does not arise.

Further, due to our design procedure for the tap positions corresponding to the variables $X_1,\ldots,X_n$, the overlap of inputs in the calls to $\sym{Maj}$ 
in the XOR of any two keystream bits is at most $\nu$. From Table~\ref{tab-nu-delta}, we observe that the value of $\nu$ is considerably smaller than $m$. So the complexity
of $\sym{Maj}$ is preserved in the XOR of any number of keystream bits.

Further, in the present version, during the initialisation phase we inject the output of the filtering function at multiple positions of the state, the number
of positions being the weight of the string $\mathbf{d}$ (see~\eqref{eqn-feed-back-cond}) which is about $\mu=\lfloor \sqrt{2\kappa}\rfloor$. Further, the positions of the 1's 
in $\mathbf{d}$ are more or less equi-spaced so that the difference between two 1-positions is also about $\mu$. 
So any (controlled) difference in any bit position travels at most $\mu$ steps before it is further modified. 
In the full initialisation phase consisting of $2\kappa$ iterations, a difference
will be updated about $\sqrt{2\kappa}$ times. This makes it difficult to control a difference through all the initialisation rounds. 

Due to the above modifications, the attack in~\cite{cryptoeprint:2025/197} on the previous proposal does not apply to the modified proposal. In fact, the modifications
though inexpensive, substantially improve the differential properties of the keystream and also improve the ``robustness'' of the initialisation phase. We note,
however, unlike our analysis of the some of the other attacks, we do not have any proof that our proposal resists all kinds of differential attacks. 
We welcome further analysis of our proposals, including finding other avenues of attack.

\section{Efficiency of Computing $\sym{MM}_n$ \label{sec-MM-eff} }
In Section~\ref{sec-crypto}, we proposed using $f_{2m+1}$ (which is $1\oplus \sym{MM}_{2m+1}$) as the filtering function in the 
nonlinear filter model. Further, Table~\ref{tab-Lm-pairs} provides
specific suggestions of values of $m$ for achieving different security levels. In this section, we consider the complexity of implementing $\sym{MM}_{2m+1}$.
Since $\sym{MM}_{2m+1}$ is constructed from $\sym{MM}_{2m}$ using one XOR gate, we need to consider the complexity of implementing $\sym{MM}_{2m}$.

The computation of $\sym{MM}_{2m}$ requires computing $\sym{Maj}_m$ and an inner product of two $m$-bit strings.
We discuss basic strategies for implementing these operations which provide estimates of the number of gates required. 

\paragraph{Inner product of two $m$-bit strings.} This operation requires $m$ AND gates and $m-1$ XOR gates.

\paragraph{Computation of $\sym{Maj}_m$.}
From Theorem~4.1 of~\cite{We87} it follows that $\sym{Maj}_m$ can be computed using $O(m)$ gates. The approach is to first compute
the weight of an $m$-bit string and then compute the majority function on the weight. We provide details of the specific approach that we followed
for these two steps.
We first consider the computation of the majority function on the weight, and then we consider the computation of the weight of an $m$-bit string.

\paragraph{Majority function on the weight of an $m$-bit string.}
We start with an example. From Table~\ref{tab-Lm-pairs}, one of the choices of $m$
is $m=37$. The weight of a 37-bit string is a 6-bit quantity, say $w_5w_4w_3w_2w_1w_0$. It is required to determine whether the value represented by this string is at least 19. 
This is computed by the Boolean formula $w_5\vee (w_4\wedge (w_3\vee w_2\vee (w_1 \wedge w_0)))$, requiring 3[OR]+2[AND] gates. 
In general, the weight of an $m$-bit string is
an $\omega$-bit value, where $\omega=\lceil\log_2m\rceil$. To compute the threshold function, $\omega_1$ OR and $\omega_2$ AND gates are required for some values 
of $\omega_1$ and $\omega_2$ satisfying $\omega_1+\omega_2\leq \omega$. 
So the number of gates for computing the threshold function on the weight of an $m$-bit string requires a logarithmic (in $m$) number of gates. 

\paragraph{Weight of an $m$-bit string.} 
A half-adder takes two input bits and outputs two bits which represent sum of the two input bits. A full adder takes three input bits and outputs two bits which represent the
sum of the three input bits. We estimate the numbers of half and full adders that are required for computing the weight of an $m$-bit string $\mathbf{x}$.
The algorithm for computing weight that we use is from~\cite{DBLP:journals/eccc/ECCC-TR05-049}. 
(Another algorithm to compute the weight is given in~\cite{DBLP:journals/ipl/DemenkovKKY10}.)
If $m=1$, then no adders are required, 
if $m=2$, a half adder computes the weight, and if $m=3$, a full adder computes the weight. For $m>3$, write $m=m_1+m_2+1$, where $m_1+1$ is the highest power of two that 
is at most $m$. The algorithm computes the weight of the first $m_1$ bits of $\mathbf{x}$, the weight of the next $m_2$ bits of $\mathbf{x}$, and then adds these two weights 
together with the last bit of $\mathbf{x}$. 

Using the above algorithm, it is easy to show that the computation of the weight of an $m$-bit string, when $m=2^r-1$ with $r\geq 1$, requires $2^r-r-1$ full adders.
We are interested in the exact counts
of full and half adders required for the values of $m$ in Table~\ref{tab-Lm-pairs}. Let us denote a full adder by [F] and a half adder by [H].
In Table~\ref{tab-Lm-pairs}, one of the choices is $m=37$. Writing $37=31+5+1$, it is required to find the weight of one $31$-bit string, one $5$-bit string and then perform
the final addition. Computation of the weight of the 31-bit string requires 26[F]. Writing $5=3+1+1$, the computation of the weight of a 5-bit string
requires 3[F] (a full adder to compute the weight of a 3-bit string, and 1[F]+1[H] to add the remaining two bits to this weight). The weight of a 5-bit string
is a 3-bit quantity, while the weight of a 31-bit string is a 5-bit quantity. So the final addition of the weights along with the remaining bit requires 
3[F]+2[H]. So a total of 31[F]+3[H] is required to compute the weight of a 37-bit string.
In a similar manner, it is possible to obtain the numbers of full and half adders required to compute the weights of $m$-bit strings for the values of $m$ given in 
Table~\ref{tab-Lm-pairs} and these counts are given below.
\begin{tabbing}
	\ \ \= $m=87$: 80[F]+3[H]; \ \ \ \ \= $m=101$: 94[F]+4[H]; \ \ \ \ \= $m=115$: 108[F]+3[H]. \kill
	\> $m=37$: 31[F]+3[H]; \> $m=59$: 53[F]+1[H]; \> $m=71$: 64[F]+3[H]; \\
	\> $m=87$: 80[F]+2[H]; \> $m=101$: 94[F]+3[H]; \> $m=115$: 108[F]+2[H].
\end{tabbing}

\paragraph{Circuit size for computing $\sym{MM}_n$.} The inner product requires $O(m)$ OR and AND gates and the weight computation requires $O(m)$ full adders.
So the circuit size for computing $\sym{MM}_n$ is $O(n)$. 

\paragraph{Concrete gate count estimates for $\mathcal{S}(L,m)$.}
Given $L$ and $m$, the gate count estimate for $\mathcal{S}(L,m)$ primarily arises from the gate count estimates of the $L$-bit state and the filtering function $f_{2m+1}$. 
Beyond these two components, a small number of XOR gates are required to compute the next bit function $\sym{nb}$ of the LFSR (see Table~\ref{tab-prim-poly}) and 
$\mu=\lfloor\sqrt{2\kappa}\rfloor$ XOR gates
are required by the initialisation round function $\sym{IR}$ (see~\eqref{eqn-IR} and the discussion following Proposition~\ref{prop-IR-inv}).
From~\eqref{eqn-f}, the estimate for the gate count of $f_{2m+1}$ is 
$2$ plus an estimate of the gate count for computing $(\sym{rev},\sym{Maj}_m)\mbox{-}\sym{MM}_{2m}$. From the above discussion, an estimate of the gate count
for $\sym{MM}_{2m}$ is $m+\lceil\log_2m\rceil$ AND/OR gates, and $m-1$ XOR gates, plus the number of gates required to compute the weight of an $m$-bit string.

To obtain concrete estimates, it is convenient to convert the various gate counts into a single unit. Previous 
works~\cite{DBLP:journals/ijwmc/AgrenHJM11,DBLP:series/lncs/CanniereP08} have taken a single NAND gate as the basic unit and translated other gates in terms of this unit. 
A half-adder can be implemented using 5 NAND gates, while a full adder can be implemented using 9 NAND gates. 
In~\cite{DBLP:journals/ijwmc/AgrenHJM11,DBLP:series/lncs/CanniereP08}, a XOR gate was taken to be 2.5 units and an AND gate was taken to be 1.5 units. 
Between the papers~\cite{DBLP:journals/ijwmc/AgrenHJM11} and~\cite{DBLP:series/lncs/CanniereP08} there is a difference in the number of units required
for a flip-flop: ~\cite{DBLP:journals/ijwmc/AgrenHJM11} takes a flip-flop to be 8 units, while~\cite{DBLP:series/lncs/CanniereP08} takes a flip-flop to be 12
units. In Table~\ref{tab-gate-cnt}, we provide estimates of the circuit sizes of $\mathcal{S}(L,m)$ for the values of $(L,m)$ in Table~\ref{tab-Lm-pairs}.
These estimates consider a flip-flop to be 8 units. 

\begin{table}
\centering
	{\scriptsize
	\begin{tabular}{|c|c|c|c|c|c|c|}
		\cline{2-7}
		\multicolumn{1}{c|}{}& 
			$\mathcal{S}(163,37)$ & $\mathcal{S}(257,59)$ & $\mathcal{S}(331,70)$ & $\mathcal{S}(389,87)$ & $\mathcal{S}(449,100)$ & $\mathcal{S}(521,114)$ \\ \hline
		LFSR        & 1304   & 2056   & 2648   & 3112   & 3592   & 4168 \\ \hline
		$f_{2m+1}$  & 453.5  & 729.5  & 888.0  & 1091.0 & 1278.0 & 1455.0 \\ \hline
		$\sym{nb}$  & 10     & 15     & 15     & 15     & 20     & 15     \\ \hline
		$\sym{IR}$  & 30     & 40     & 45     & 47.5   & 52.5   & 55     \\ \hline
		total       & 1797.5 & 2840.5 & 3596.0 & 4265.5 & 4942.5 & 5693.0 \\ \hline
	\end{tabular}
	\caption{Estimates of the number of gate units required to implement $\mathcal{S}(L,m)$ for values of $L$ and $m$ in Table~\ref{tab-Lm-pairs}. 
Following~\cite{DBLP:journals/ijwmc/AgrenHJM11,DBLP:series/lncs/CanniereP08}, a single NAND gate is taken to be a basic unit, a XOR gate to be 2.5 units and an
AND gate to be 1.5 units. Following~\cite{DBLP:journals/ijwmc/AgrenHJM11} a flip-flop is taken to be 8 units. 
	\label{tab-gate-cnt} }
	}
\end{table}

If we consider a flip-flop to be 12 units as done for Trivium~\cite{DBLP:series/lncs/CanniereP08}, then the gate count estimate for $\mathcal{S}(163,37)$ is
2404.5. The gate estimate for Trivium obtained in~\cite{DBLP:series/lncs/CanniereP08} is 3488. Both Trivium and $\mathcal{S}(163,37)$ are targeted at the
80-bit security level, and $\mathcal{S}(163,37)$ is substantially smaller than Trivium. The lower size of $\mathcal{S}(163,37)$ is due to the smaller
state size; $\mathcal{S}(163,37)$ uses a 163-bit state, while Trivium uses a 288-bit state.

The gate count estimate for Grain-128a obtained in~\cite{DBLP:journals/ijwmc/AgrenHJM11} is 2145.5. Grain-128a is targeted at the 128-bit security level.
Comparing to $\mathcal{S}(257,59)$, which is also targeted at the 128-bit security level, we find the gate count estimate of $\mathcal{S}(257,59)$
to be 2840.5. The state sizes of both Grain-128a and $\mathcal{S}(257,59)$ are almost the same. The greater size of $\mathcal{S}(257,59)$ is due to the
greater gate size requirement of the filtering function $f_{2m+1}$ in comparison to the gate size requirement of the nonlinear components of Grain-128a. 
Even though $\mathcal{S}(257,59)$ is larger than Grain-128a, its size of about
2840.5 gates is small enough for $\mathcal{S}(257,59)$ to be considered as a small cipher. Finally we note that $\mathcal{S}(521,114)$ which is targeted
at the 256-bit security level requires about 5693.0 gates. We know of no other 256-bit secure stream cipher which has such a small gate count. 

We note that it is possible to improve security at the cost of a reasonable increase in gate count. For example, from Table~\ref{tab-Lm-pairs},
at the 128-bit security level, our proposal has $L=257$ and $m=59$, resulting in linear bias equal to $2^{-60}$ and the fast algebraic attack requiring more than
$2^{130.12}$ operations. If we increase $m$ from 59 to 63, then the linear bias drops to $2^{-64}$, and the fast algebraic attack now requires more than
$2^{135.83}$ operations. The gate estimate for $\mathcal{S}(257,63)$, i.e. the stream cipher with $L=257$ and $m=63$, is 2887.5 gates (the LFSR requires
2056 gates, and $f_{127}$ requires 776.5 gates). So the security increases by about 5 bits at the cost of an increase of only 47 gates in the circuit size.


\section{Conclusion \label{sec-conclu} }
We described a construction of balanced Boolean functions with several provable properties, namely very high nonlinearity, acceptable algebraic resistance, 
\textit{and} enabling efficient hardware implementation. Using such Boolean functions, we proposed concrete constructions of the nonlinear filter model for stream ciphers targeted
at different security levels. Gate count estimates for the stream cipher proposals show that the circuit sizes compare well with famous stream ciphers at the 80-bit
and the 128-bit security levels, while for higher security levels, we do not know of any stream cipher with lower gate count estimates.

\section*{Acknowledgement} 
We thank Subhadeep Banik, Tim Beyne, Pierrick M\'{e}aux, Willi Meier, Michiel Verbauwhede, and Bin Zhang for their comments on earlier versions of the paper. 
We also thank the anonymous reviewers for helpful comments.
Deng Tang provided us with a program written by Simon Fischer which we have used for computing fast algebraic immunity. We thank both of them.


\end{document}